\documentclass[12pt,a4paper]{article}
\usepackage[utf8]{inputenc}
\usepackage{a4wide}
\usepackage{graphicx}
\usepackage{amsmath}
\allowdisplaybreaks[1] 
\usepackage{hyperref}
\usepackage{cancel}
\usepackage{cite}
\usepackage{multirow}
\usepackage{slashed}
\usepackage{amssymb}

\usepackage[dvipsnames]{xcolor}

\usepackage{pifont}
\newcommand{\red}[1]{{\color{red} #1}}
\newcommand{\green}[1]{{\color{Green} #1}}

\newcommand{\newcheckmark}{\green{\textrm{\ding{52}}}}%
\newcommand{\newcrossmark}{\red{\textrm{\ding{56}}}}%

\begin{document}

\begin{titlepage}
\begin{flushright}
LU-TP 22-63\\
November 2022
\end{flushright}
\vfill
\begin{center}
{\Large\bf Constraints on the hadronic light-by-light in the Melnikov-Vainshtein regime
}
\vfill
{\bf Johan Bijnens $^{a,\, \dagger}$, Nils Hermansson-Truedsson $^{a,\, b, \, \ddagger}$, Antonio Rodr\'{i}guez-S\'{a}nchez $^{c,d, \, \ast}$}\\[0.3cm]
{$^{a}$ Department of Astronomy and Theoretical Physics, Lund University,
\\
Box 43, SE 221 00 Lund, Sweden
\\
$^{b}$ Albert Einstein Center for Fundamental Physics, Institute for Theoretical Physics, Universit\"{a}t Bern, Sidlerstrasse 5, 3012 Bern, Switzerland
\\
$^{c}$ Scuola Internazionale Superiore di Studi Avanzati (SISSA), 
\\
via Bonomea, 265 - 34136 Trieste, Italy
\\
$^{d}$ INFN, Sezione di Trieste, SISSA, Via Bonomea 265, 34136, Trieste, Italy
}
\end{center}
\vfill
\begin{abstract}
The muon anomalous magnetic moment continues to attract attention due to the possible tension between the experimentally measured value and the theoretical Standard Model prediction. With the aim to reduce the uncertainty on the hadronic light-by-light contribution to the magnetic moment, we derive short-distance constraints in the Melnikov-Vainshtein regime which are useful for data-driven determinations. In this kinematical region, two of the four electromagnetic currents are close in the four-point function defining the hadronic light-by-light tensor. To obtain the constraints, we develop a systematic operator product expansion of the tensor in question to next-to-leading order in the expansion in operators. We evaluate the leading in $\alpha_s$ contributions and derive constraints for the next-to-leading operators that are also valid nonperturbatively.
\end{abstract}
\vfill

{\footnotesize\noindent $^\dagger$ johan.bijnens@thep.lu.se
\\
\footnotesize\noindent $^\ddagger$ nils.hermansson-truedsson@thep.lu.se
\\
\footnotesize\noindent $^\ast$ arodrigu@sissa.it
}
\end{titlepage}

\tableofcontents
\section{Introduction}
\label{sec:introduction}

\begin{figure}[t!]
	\centering
	\includegraphics[height=0.17\textheight]{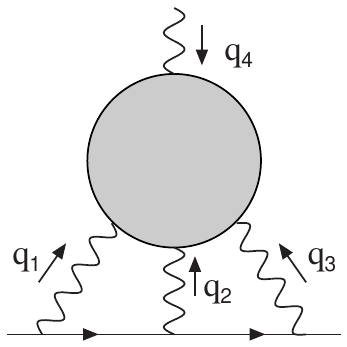}
	\caption{ The HLbL contribution to $a_\mu$. The grey blob contains hadronic contributions. The photon with momentum $q_4$ is the external magnetic field. The bottom line is the muon and the wiggly lines are photons.}\label{fig:hlbl}
\end{figure}

The object of study in this work is the hadronic light-by-light (HLbL) tensor\footnote{The integral over $q_4$ simply removes the delta function for conservation of momenta.},
\begin{equation}\label{eq:hlbltensor}
\Pi^{\mu_{1}\mu_{2}\mu_{3}\mu_{4} } 
=
-i\int \frac{d^{4}q_{4}}{(2\pi)^{4}}\left(\prod_{i=1}^{4}\int d^{4}x_{i}\, e^{-i q_{i} x_{i}}\right)  \langle 0 | T\left(\prod_{j=1}^{4}J^{\mu_{j}}(x_{j})\right)|0\rangle \, .
\end{equation}
Above, the electromagnetic current is defined as
\begin{equation}
J^{\mu}(x)={\sum_{q}}e_q\, \bar{q}(x)\gamma^{\mu}q(x)  \, ,
\end{equation}
with $q=u,d,s$ and $e_q$ the associated light-quark charge matrix,  $e_q=\frac{1}{3}\left(2,-1,-1\right)$.
In spite of its very simple definition, only involving color symmetry and three light quarks, its emerging dynamics is extremely complex and rich, as a consequence of the underlying nonperturbative color dynamics. Its multi-scale nature represents a theoretical challenge to understand where and how a picture in terms of approximately free quarks and gluons can be used to describe it, and how to separate the corresponding asymptotic QCD dynamics from the nonperturbative hadronic one. Achieving such a task with a systematic and unambiguous power counting, employing rigorous OPE methods, is one of the motivations of this research line, started in Refs.~\cite{Bijnens:2019ghy,Bijnens:2020xnl,Bijnens:2021jqo} and continued here in a different kinematic regime, all of them in the region where one of the external photons becomes soft. The motivation to focus on that region is the key phenomenological role played by the HLbL tensor in theoretically assessing the anomalous magnetic moment of the muon via the diagram depicted in Fig.~\ref{fig:hlbl}. Indeed, its contribution represents one of the main uncertainties in the most recent Standard Model (SM) value given in the White Paper~\cite{Aoyama:2020ynm}, 
\begin{equation}
a_{\mu}^{\textrm{WP}}   =  \frac{(g-2)_\mu}{2}= 116 \, 591 \, 810(43) \times 10^{-11}\, .
\end{equation}
The quoted value is $4.2 \sigma$ away from the experimental one \cite{Muong-2:2006rrc,Muong-2:2021ojo},
\begin{equation}
    a_\mu ^{\textrm{exp}}= 116 \,  592\,  061(41) \times 10^{-11} \, .
\end{equation}
However, a recent lattice evaluation~\cite{Borsanyi:2020mff} of the leading hadronic vacuum polarization contribution (HVP) suggests that $R$-ratio based evaluations might be underestimating its contributions, possibly explaining the tension without any need of Beyond the Standard Model (BSM). A short discussion and more relevant references can be found in Ref.~\cite{Colangelo:2022jxc}.

While clarifying the situation with respect to the HVP side has become the most urgent matter, further improvements on the HLbL evaluation are needed in order to match the improving experimental precision. The first proper evaluations date back to 1990s \cite{Hayakawa:1997rq,Hayakawa:2001bb,Bijnens:1995xf,Bijnens:2001cq} and were mainly based on models. 
Our study is part of the data-driven evaluation of HLbL, which is based on dispersion relations for the leading low-energy contributions~\cite{Colangelo:2015ama,Colangelo:2017fiz}. The subleading, but not negligible, contributions from the short- and intermediate-distance regions, i.e., involving any large or intermediate loop momenta, are quite challenging to assess within the dispersive approach, and one needs to rely on models and other constraints. A thorough discussion can be found in the White Paper \cite{Aoyama:2020ynm}. Here we simply quote the dispersive White Paper prediction~\cite{Aoyama:2020ynm}
\begin{align}
    a_{\mu}^{\textrm{HLbL}} = 92(19)\times 10^{-11} \, ,
\end{align}
which is based on Refs.~\cite{Melnikov:2003xd,Masjuan:2017tvw,Colangelo:2017fiz,Hoferichter:2018kwz,Gerardin:2019vio,Bijnens:2019ghy,Colangelo:2019uex,Pauk:2014rta,Danilkin:2016hnh,Jegerlehner:2017gek,Knecht:2018sci,Eichmann:2019bqf,Roig:2019reh}.\footnote{Let us also point out that major improvements on the lattice-based evaluation have been made since them. See Ref.~\cite{Chao:2021tvp}.}

Systematic uncertainties from the short-distance (SD) behaviour of the HLbL tensor can be controlled with the help of short-distance constraints (SDCs). Such nontrivial information about the analytical structure of the HLbL tensor can be obtained through rigorous
Operator Product Expansion (OPE) methods. These OPEs are constructed for different limits of the Euclidean virtualities $Q_{i}^2=-q_{i}^2$ associated to the virtual photons in Fig.~\ref{fig:hlbl}. 
In Refs. \cite{Bijnens:2019ghy,Bijnens:2020xnl} we showed how a systematic OPE \cite{Balitsky:1983xk,Ioffe:1983ju}, different from the vacuum OPE,  which is valid when all Euclidean loop momenta of the HLbL tensor are large~\cite{Shifman:1978bx}, can be applied to the HLbL when all the loop momenta in Fig.~\ref{fig:hlbl} are large, i.e.~$Q_{i}^2\gg \Lambda _{\textrm{QCD}}^2$, in spite of the soft nature ($q_4\rightarrow 0$) of the photon coupling to the external magnetic field. The naive massless quark loop happened to be the leading contribution, directly related to the fact that 
it comes from the photon operator $F_{\mu\nu}$, that, up to higher orders in QED, is not affected by quark-gluon interactions. The perturbative corrections to the Wilson coefficient of this operator were calculated in Ref.~\cite{Bijnens:2021jqo}.

Another regime of short-distances for the HLbL tensor and its contribution to the muon $g-2$ is where two of the Euclidean momenta are large. The study of this area was pioneered by Ref.~\cite{Melnikov:2003xd}. In this limit a second Euclidean loop momentum, apart from the static $q_4 \rightarrow 0$ photon corresponding to the magnetic field, becomes much smaller than the other two. The leading term and some of the implications were worked out in Ref. \cite{Melnikov:2003xd}. While there is some debate about the phenomenological consequences of the corresponding SD constraints \cite{Melnikov:2019xkq,Colangelo:2019uex,Colangelo:2019lpu}, their validity is well-established and relevant for constraining the HLbL, see e.g.~Refs.~\cite{Colangelo:2019lpu,Colangelo:2019uex,Melnikov:2019xkq,Leutgeb:2019gbz,Ludtke:2020moa,Knecht:2020xyr,Masjuan:2020jsf,Cappiello:2021vzi,Colangelo:2021nkr,Danilkin:2021icn,Leutgeb:2021bpo,Leutgeb:2021mpu,Zanke:2021wiq} for implementations of the SD constraints for various models and intermediate contibutions. The White Paper estimate for the short-distance contribution to the HLbL is~\cite{Aoyama:2020ynm}
\begin{align}
    \Delta a_\mu^{\textrm{HLbL},\, \textrm{SDC}} = 15(10) \times 10^{-11} \, .
\end{align}

The main goal of this work consists in developing the associated OPE beyond the limits obtained in \cite{Melnikov:2003xd}, adding power corrections and parameterizing the nonperturbative ignorance in a few form factors whose energy dependence are known for large momenta. 

In Sec.~\ref{sec:generalities} we present some more details on the general properties of the HLbL. The OPE studied in this work is formulated in Sec.~\ref{sec:OPE} through next-to-leading order (NLO). In Sec.~\ref{sec:alphaSLO} we focus on the region where the matrix elements can be computed perturbatively up to NLO in the energy expansion and compare with the results of Ref. \cite{Bijnens:2021jqo}. 
The full agreement with the corresponding expansion of the OPE of Ref.~\cite{Bijnens:2021jqo} is also shown in that section, leading to a very strong consistency check of the interplay of the different two regions. Finally in Sec.~\ref{sec:nonperturbative} we perform a nonperturbative extrapolation at leading order in $\alpha_s$ at the hard scale, parametrizing the nonperturbative effects in a few scalar functions depending on only a single squared momentum. Conclusions are then given in Sec.~\ref{sec:conclusions}.

\section{Generalities of the hadronic light-by-light}
\label{sec:generalities}

The HLbL contribution to the muon magnetic moment, $a_\mu^{\textrm{HLbL}}$, is depicted diagrammatically in Fig.~\ref{fig:hlbl}. 
The underlying object to consider is thus a four-point function of electromagnetic currents, i.e.~the HLbL tensor as defined in Eq.~(\ref{eq:hlbltensor}). We denote the virtual photon momenta by $q_{1}$, $q_2$, $q_3$, and the external photon momentum $q_4$ which is soft for the $(g-2)_\mu$ kinematics, it corresponds to the external magnetic field. These momenta are chosen to satisfy $q_1+q_2+q_3+q_4=0$.
We denote the magnitudes of the three momenta in the Euclidean by $Q_i$, i.e. $q_i^2=-Q_i^2$. The tensor in~(\ref{eq:hlbltensor}) satisfies the Ward identities
\begin{align}
    q_{i,\, \mu _i} \, \Pi ^{\mu _1 \mu_2 \mu_3 \mu_4} &= 0 \, , \qquad i=1,\dots,4\, ,
\end{align}
which immediately leads to
\begin{align}\label{eq:wardcons}
\Pi^{\mu_{1}\mu_{2}\mu_{3}\mu_{4}}=-q_{4,\, \nu_{4}}\frac{\partial \Pi^{\mu_{1}\mu_{2}\mu_{3}\nu_{4}}}{\partial q_{4,\, \mu_{4}}}\, .
\end{align}
The above relation implies that the derivative of the HLbL tensor at $q_4=0$ can be used to fully determine the HLbL contribution to the magnetic moment, $a_\mu^{\textrm{HLbL}}$~\cite{Aldins:1970id}. 

For this reason, we will in the following discuss the derivative of the HLbL tensor
\begin{equation}
\lim_{q_4\rightarrow 0} \frac{\partial \Pi^{\mu_1\mu_2\mu_3\nu_4}}{\partial q_{4,\mu_4}} \, .
\end{equation}
This tensor can be decomposed into a set of $19$ independent scalar form factors at $q_4\to0$ \cite{Melnikov:2003xd,Colangelo:2015ama}. However, these may be generated from only $5$ different form factors $\hat\Pi _j$ by using crossing relations~\cite{Colangelo:2015ama}. The precise relation to the derivative can be found in Ref.~\cite{Colangelo:2015ama}.

As explained in detail in Refs.~\cite{Colangelo:2015ama,Colangelo:2017fiz}, from the five independent $\hat\Pi_j$ functions one can construct twelve scalar functions $\overline\Pi _i$ such that
 $a_\mu^{\textrm{HLbL}}$ is given by
\begin{align}\label{eq:amuint}
a_{\mu}^{\textrm{HLbL}} 
= 
\frac{2\alpha ^{3}}{3\pi ^{2}} 
& \int _{0}^{\infty} dQ_{1}\int_{0}^{\infty} dQ_{2} \int _{-1}^{1}d\tau \, \sqrt{1-\tau ^{2}}\, Q_{1}^{3}Q_{2}^{3}
\sum _{i=1}^{12} T_{i}(Q_{1},Q_{2},\tau)\, \overline{\Pi}_{i}(Q_{1},Q_{2},\tau)\, ,
\end{align}
where the $T_i $ are known kernel functions and $\tau$ is defined by $Q_3^2=Q_1^2+Q_2^2+2\tau Q_1Q_2$.

\section{The two-current OPE}
\label{sec:OPE}

The HLbL tensor in Eq.~(\ref{eq:hlbltensor}) may be rewritten as 
\begin{align}\label{eq:twopoint}
\Pi^{\mu_{1}\mu_{2}\mu_3\mu_4}=
&
\sum_{j,k}\frac{i e_{q_j}e_{q_k}}{e^{2}}\int \frac{d^{4}q_4}{(2\pi)^4}\int d^{4}x_1\int d^{4}x_2\, e^{-i(q_1 x_1+q_2 x_2)}
\nonumber \\
& \times 
\langle 0 |T(J_j^{\mu_1}(x_1)J_k^{\mu_2}(x_2))|\gamma^{\mu_3}(q_3)\gamma^{\mu_4}(q_4) \rangle \, ,
\end{align}
where we introduced a flavor-index on the current such that
\begin{equation}
J^\mu_j=\bar{q}_j\gamma^\mu q_j\, .    
\end{equation}
Note that two of the photons have been captured in the external state, which can be contracted with electromagnetic interaction terms in the implicit Dyson series with the QCD+QED action. The Lorentz indices on the photons in the external state is a convenient short-hand notation for contraction with a photon field $A^\alpha$ in the Dyson series according to 
\begin{equation}
A^{\alpha}(x)|\gamma^{\mu}(q)\rangle\equiv g^{\alpha\mu}e^{-iqx} \, .
\end{equation}
The difference to the usual contraction
\begin{equation}
A^{\alpha}(x)|\gamma(q)\rangle= \epsilon^{\alpha}(q)e^{-iqx} \, ,
\end{equation}
is that we factor out the photon polarization vector $\epsilon ^{\alpha}(q)$. 
The regime we study in this paper is where we can perform a proper OPE on the time ordered product of two quark currents, $T(J_j^{\mu_1}(x_1)J_k^{\mu_2}(x_2))$. From the operators appearing in this OPE we only need to keep those that can mediate the transition to two extra photons, one of which, $\gamma^{\mu_4}(q_4)$, is in the static $q_4\rightarrow 0$ limit relevant for the $(g-2)_\mu$.

The OPE is valid as far as there is an energy scale flowing
between both space-time points much larger than the rest of scales entering into the process, which can then be factored out from the long-distance problem. Often the mass of a heavy propagator plays such a role and we refer to the corresponding OPE application as integrating out a heavy particle. In this case the propagators, as far as the quark-gluon picture holds, are (almost) massless. Nevertheless, in the kinematic regime in which most of the momentum $q_1$ (which must be large enough with respect to the dynamically generated $\Lambda_{\mathrm{QCD}}$ scale, in such a way that the quark-gluon picture holds) that enters at $x_1$, leaves the process at $x_2$, an OPE can be performed, taking $x_1$ and $x_2$ close to each other. We thus define $\hat q$ via
\begin{align}
\hat q &= \frac{1}{2}\left(q_1-q_2\right)\, ,& q_{1,2} = \pm\hat q-\frac{1}{2}\left(q_3+q_4\right)\,.
\end{align}
The regime we consider is taking $q_3$ and $q_4$ much smaller than $\hat q$. The large scale is then given by $\hat Q=\sqrt{-\hat q^2}$. Of course, in the integration over loop momenta in Eq.~(\ref{eq:amuint}) one has to consider also the cases with $q_1$ or $q_2$ small, but these can be handled analogously. 

When dealing with the contribution to $a_\mu^{\textrm{HLbL}}$ more convenient variables are $\overline Q_3=Q_1+Q_2$, $\delta_{12}=Q_1-Q_2$ and $Q_3$. The sizes of these variables are constrained, when $q_4\to0$, by $q_1+q_2+q_3=0$ implying $\vert \delta_{12}\vert\le Q_3$.  The large $\hat Q$ and the large $\overline Q_3$ expansions are related through the identity
\begin{align}\label{eq:qhatrelation}
    \hat{Q}^2 = \frac{1}{4} \left( \overline{Q}_3 ^2+ \delta _{12}^2 -Q_{3}^2 \right) \, .
\end{align}
However, we stress that there are three variables to expand in here, one large and two small. 
In the large $\hat{Q}$ case, the small variables are $q_3/\hat Q$ and $q_3\cdot \hat{q}/\hat Q^2$. For large $\overline Q_3$ it is instead $Q_3/\overline Q_3$ and $\delta _{12}/\overline Q_3$ that are small. We will use the $\hat Q$ large expansion for the OPE and the large $\overline Q_3$ expansion for the final expressions for the $\hat\Pi_i$. We find the latter, also used in the so-called corner region in  Ref.~\cite{Bijnens:2021jqo}, more convenient for the $a_\mu^{\textrm{HLbL}}$ integrals in Eq.~(\ref{eq:amuint}), since the mapping to the Euclidean variables $Q_1, Q_2, Q_3$ is simpler. Nevertheless, since we are working at NLO in the $\frac{1}{\overline{Q}_3}$ expansion, one can trivially identify $\overline{Q}_3=2\, \hat{Q}$ for the final $\hat{\Pi}_i$ result, keeping $\delta_{12}$ and $Q_{3}$ small. It should be noted that the LO $\hat{\Pi}_i$ results were calculated in Ref.\cite{Colangelo:2019uex} for the $q_3$ small region. With the identification $\overline{Q}_3=2\, \hat{Q}$, the NLO case becomes straightforward.\footnote{The $\hat{\Pi}_i$ for $i\neq 1,4,7,17,39,54$ given in Ref.~\cite{Colangelo:2019uex} in the $q_3$-small region can be trivially recovered by using our results for $q_1$ and $q_2$ small and then applying the crossing relations given in Eq.~(2.8) of Ref.\cite{Colangelo:2019uex}. For example $\hat{\Pi}_{10}(Q_1,Q_2,Q_3)=\hat{\Pi}_{7}(Q_1,Q_3,Q_2)$. Then, recovering the $\hat{\Pi}_{10}$ at small $Q_3$ is straightforward from our expression at small $Q_2$. In that specific case in the perturbative regime one finds from Eq.~(\ref{eq:d4pihatfinalq2}), 
\begin{equation}
\hat{\Pi}_{10}=-N_c\sum_q e_q^4\frac{16}{3\pi^2 Q_3^2\overline{Q}_3^4}=-\frac{2}{3}\frac{1}{16}\frac{16}{3\pi^2 Q_3^2 \hat{Q}^4}=\frac{2}{9\pi^2q_3^2 \hat{q}^4} \, .
\end{equation}
}

\begin{figure}
    \centering
    \includegraphics{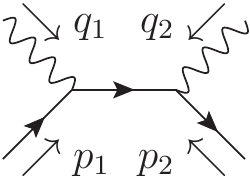}
    \caption{The diagrams needed for the lowest order in $\alpha_s$ OPE. The diagram with $q_1$ and $q_2$ interchanged must be added.
    }
    \label{fig:OPELO}
\end{figure}

Since the object in Eq.~(\ref{eq:twopoint}) has dimension $-2$ in energy, a local operator of dimension $D$ is going to give a contribution suppressed by $\frac{1}{\overline{Q}_3^{D-2}}$. Let us now perform the derivation of the OPE up to $D=4$. We may start by the contributions given by operators involving quark fields. Let us first simplify the notation of the integration and matrix element in Eq.~(\ref{eq:twopoint}). To do this, we note that at leading order in $\alpha_s(\overline{Q}_3)$ one has to contract two of the quark fields in the two electromagnetic currents in the product $J_j^{\mu_1}(x_1)J_k^{\mu_2}(x_2)$, thus giving rise to a quark propagator and the restriction $j=k$. The only way of achieving a potential $j \neq k$ contribution at NLO would be with a gluon propagator between the two quark lines. In order to get a local contribution of $D\leq 4$ (terms of increasing dimension in the OPE are suppressed by larger powers of $\frac{Q_3,\Lambda_{\mathrm{QCD}}}{\overline{Q}_3}$) one needs to close at least one extra quark line, but then the associated color trace vanishes. Contributions with $j \neq k$ in this regime can then only start at $\mathcal{O}\left(\alpha_s^2(\overline{Q}_3)\right)$ (and only through $D=4$ field strength contributions, because of charge conjugation) and can thus be neglected. We may thus write Eq.~(\ref{eq:twopoint}) as
\begin{align}
\Pi^{\mu_{1}\mu_{2}\mu_3\mu_4}&=
\sum_{j}\frac{i e_{q_j}^2}{e^{2}}\int \frac{d^{4}q_4}{(2\pi)^4}\int d^{4}x_1\int d^{4}x_2\, e^{-i(q_1 x_1+q_2 x_2)}
\nonumber \\
& \qquad \qquad 
\times
\langle 0 |T(J_j^{\mu_1}(x_1)J_j^{\mu_2}(x_2))|\gamma^{\mu_3}(q_3)\gamma^{\mu_4}(q_4) \rangle 
\nonumber \\
&\equiv
\sum_j \frac{i e_{q_j}^{2}}{e^2} \Bigg\langle e^{-i(q_1 x_1 +q_2 x_2)}
\, \bar{q}(x_1)\gamma^{\mu_1}q(x_1) \;\bar{q}(x_2)\gamma^{\mu_2}q(x_2) \Bigg\rangle ^{j,\mu_3,\, \mu_4}_{q_{4},\, x _1,\, x_2} \label{eq:tensorbracket}
\, .
\end{align}
In this notation the subscripts on the angular bracket on the right refers to integration variables, $j$ in the superscript to the specific quark-flavor in the summation, and $\mu_{3},\mu_4$ indicate the matrix element with photons in the final state. Defining the massless quark propagator $S(x)$ and its Fourier transform $S(k)$ through
\begin{equation}
S(x) = \int \frac{d^4k}{(2\pi)^4} \, e^{-i k\cdot x} \, S(k) = \int \frac{d^4k}{(2\pi)^4} \, e^{-i k\cdot x} \, \frac{\slashed{k}}{k^2}\, ,
\end{equation}
we may contract two of the quark lines in Eq.~(\ref{eq:twopoint}) to yield
\begin{align}\nonumber
& \Pi^{\mu_{1}\mu_{2}\mu_3\mu_4}
=
\\ \nonumber &
=
\sum_{j}\frac{i e_{q_j}^2}{e^{2}} 
\Bigg\langle ie^{-i(q_1 x_1 +q_2 x_2)}\Big[\bar{q}(x_1)\gamma^{\mu_1}S(x_1-x_2)\gamma^{\mu_2}q(x_2)
+\bar{q}(x_2)\gamma^{\mu_2}S(x_2-x_1)\gamma^{\mu_1}q(x_1)\Big] \Bigg\rangle ^{j, \mu_3,\, \mu_4}_{q_{4},\, x _1,\, x_2}
\nonumber \\ 
&
=  \sum_{j}\frac{i e_{q_j}^2}{e^{2}}
\Bigg\langle  i e^{-i[x_1(k+q_1)+x_2(-k+q_2)]} \Big[\bar{q}(x_1)\gamma^{\mu_1}S(k)\gamma^{\mu_2}q(x_2)
-\bar{q}(x_2)\gamma^{\mu_2}S(k)\gamma^{\mu_1}q(x_1)\Big] \Bigg\rangle ^{j, \mu_3,\, \mu_4}_{q_{4},\, k ,\,  x _1,\, x_2}\, .
\end{align}
The two terms in the angular brackets take the form $\bar{q}(x_1)\Gamma \, q(x_2)$ where $\Gamma$ is a Dirac matrix. We next use translational invariance to transform to the symmetric point $(x_1-x_2)/2$ according to
\begin{align}\label{eq:transinv}
   \langle 0 | \, \bar{q}(x_1)\Gamma \, q(x_2) \,  |\gamma ^{\mu _3}(q_3)\gamma ^{\mu _4}(q_4) \rangle
   = &\, 
   \langle 0 | \, e^{i\hat{P}\cdot x_1} \bar{q}(0)\Gamma \, e^{i\hat{P}\cdot (x_2-x_1)} q(0) e^{-i\hat{P}\cdot x_2} \,  |\gamma(q_3) ^{\mu _3} \gamma ^{\mu _4}(q_4) \rangle
   \nonumber\\
   = &\, 
  e^{-i(q_3+q_4)\frac{x_1+x_2}{2}} \langle 0 | \, \bar{q}\left(\frac{x_1-x_2}{2} \right)\Gamma \,  q\left(\frac{x_2-x_1}{2}\right)\,  |\gamma(q_3)\gamma(q_4) \rangle \, ,
\end{align}
where $\hat{P}$ is a momentum operator. The benefit of this translation is that it allows to avoid total derivative operators in the OPE. At this stage we can perform the OPE for $(x_1-x_2)/2$ small. Through dimension $D=4$ one has
\begin{align}\nonumber
\Pi^{\mu_{1}\mu_{2}\mu_3\mu_4}&=-\sum_{j}\frac{ e_{q_j}^2}{e^{2}} \Bigg\langle\bar{q}(0)
\left[ \Gamma^{\mu_1\mu_2}(-\hat{q})-\Gamma^{\mu_2\mu_1}(-\hat{q})\right]
q(0) \Bigg\rangle^{j,\mu_3,\mu_4} 
\\
&
-\sum_{j}\frac{ e_{q_j}^2}{e^{2}} \Bigg\langle 
e^{-i x_1 \left[k+q_1-\frac{q_3+q_4}{2} \right] } \, 
e^{-i x_2 \left[ -k+q_2-\frac{q_3+q_4}{2} \right] }
\nonumber \\
&
\qquad \qquad \times
\frac{(x_2-x_1)_\alpha}{2} \, \bar{q}(0) 
\left[\overrightarrow{\partial}^{\alpha}-\overleftarrow{\partial}^{\alpha}\right]
\left[
\Gamma^{\mu_1\mu_2}(k)+\Gamma^{\mu_2\mu_1}(k)
\right]  
q(0) \Bigg\rangle ^{j,\mu_3,\mu_4}_{q_4,x_1,x_2,k} 
\nonumber
\, \\ \nonumber
&
= 
-\sum_{j}\frac{ e_{q_j}^2}{e^{2}}\,   \Bigg\langle\bar{q}(0)
\left[\Gamma^{\mu_1\mu_2}(-\hat{q})-\Gamma^{\mu_2\mu_1}(-\hat{q})\right]
q(0) \Bigg\rangle^{j,\mu_3,\mu_4} 
\\
&
+\sum_{j}\frac{ e_{q_j}^2}{e^{2}}\, 
\lim _{p_A \to 0 }\left(\frac{-i\, \partial ^{p_A}_{ \alpha}}{2}\right)
\Bigg\langle \bar{q}(0)
\left[
\overrightarrow{\partial}^{\alpha}-\overleftarrow{\partial}^{\alpha}
\right]
\big[
\Gamma^{\mu_1\mu_2}(\hat{q}+p_A)
\nonumber \\
& 
\hspace{150pt}
+\Gamma^{\mu_2\mu_1}(\hat{q}+p_A)
\big]  q(0) \Bigg\rangle ^{j,\mu_3,\mu_4}
\, ,
\end{align}
where $\Gamma^{\mu_1\mu_2}(k)=\gamma^{\mu_1}S(k)\gamma^{\mu_2}$ and we introduced a spurious momentum $p_A$ to remove the $x_{i,\, \alpha}$ on the second line through $x_{i, \, \alpha} = -i \lim _{p_A \to 0} \partial _{p_A , \, \alpha} e^{ix_{i}p_A}$. For $D=4$ contributions we will work in the chiral limit. Using the relation
\begin{equation}
\Gamma^{\mu_1\mu_2}(k)+\Gamma^{\mu_2\mu_1}(k)=\frac{2\, k_{\delta}\gamma_{\beta}}{k^2}(g^{\mu_1\delta}g^{\mu_2\beta}+g^{\mu_2\delta}g^{\mu_1\beta}-g^{\mu_1\mu_2}g^{\delta\beta}) \, .
\end{equation}
one finds
\begin{align}\label{eq:oped4quark}
\nonumber
\Pi^{\mu_{1}\mu_{2}\mu_3\mu_4}
&\approx - \sum_j\frac{ e_{q_j}^2}{e^2}\,   \Bigg\langle\bar{q}(0)\left[
\Gamma^{\mu_1\mu_2}(-\hat{q})-\Gamma^{\mu_2\mu_1}(-\hat{q})
\right]
q(0) \Bigg\rangle^{j, \mu_3,\mu_4} 
\\&-\sum_j\frac{ie_{q_j}^2}{e^2\hat{q}^2}\, \left(
g^{\mu_1}_{\delta}g^{\mu_2}_{\beta}+g^{\mu_2}_{\delta}g^{\mu_1}_{\beta}-g^{\mu_1\mu_2}g_{\delta\beta}\right)
\left(g_{\alpha}^{\delta}-2\, \frac{\hat{q}^\delta \hat{q}_\alpha}{\hat{q}^2}\right) 
\nonumber \\
&
\qquad \qquad \times
\Bigg\langle \bar{q}(0)\left[\overrightarrow{D}^{\alpha}-\overleftarrow{D}^{\alpha}\right] \gamma^{\beta}  q(0) \Bigg\rangle ^{j, \mu_3,\, \mu_4} \, .
\end{align}
In the last line we have promoted the partial derivative to a covariant one, which at the studied order is equivalent. Notice that the $D=4$ contribution on the second and third lines already is a power correction of the $D=3$ one, which at the same time is in a sub-leading kinematic regime for $a_\mu ^{\textrm{HLbL}}$.
Imposing gauge invariance in both sides of the equality, one can factor out $q_4$ to yield
\begin{align}\label{eq:factoriseq4}
\nonumber
\Pi^{\mu_{1}\mu_{2}\mu_3\mu_4}&=-q_{4\nu_4}\frac{\partial \Pi^{\mu_1\mu_2\mu_3\nu_4}}{\partial q_{4\mu_4}} \, , \\
\langle Q\rangle^{j,\mu_3\mu_4}&=-q_{4\nu_4}\partial^{\mu_4}_{q_4}\langle Q\rangle^{j,\mu_3\nu_4} \, ,
\end{align}
where $Q$ is one of the operators in the matrix elements in Eq.~(\ref{eq:oped4quark}). The tensor remaining after the factorisation has a well defined $q_{4}\rightarrow 0$ limit, and is antisymmetric under $\mu_4\leftrightarrow\nu_4$ exchange. Since we have factored out the heavy scale $\hat{q}$, the remaining matrix elements in the soft limit,
\begin{equation}
\lim_{q_4 \rightarrow 0}\partial^{\nu_4}_{q_4}\langle Q\rangle^{j,\mu_3\mu_4} \, ,
\end{equation}
can only depend on $q_3$. An alternative way of looking at this result is to take the diagrams of Fig.~\ref{fig:OPELO} and expand in $p_1,p_2$ small with $q_{1,2}=\pm\hat q-(1/2)\left(p_1+p_2\right)$.

\begin{figure}
    \centering
    \includegraphics{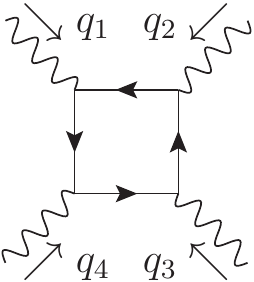}
    \caption{The diagrams needed for the lowest order in $\alpha_s$ OPE for the operators with field strength tensors. The diagrams are with the $q_3,q_4$ legs photon or gluon and all permutations of the external legs must be taken into account.}
    \label{fig:OPEFIELDLO}
\end{figure}

A second kind of $D=4$ contribution arises from field strength tensors, which are hidden in the implicit Dyson series of Eq.~(\ref{eq:transinv}). Denoting the interaction part of the action by $S_{\textrm
int}$, one has a quadratic term
\begin{equation}
e^{iS_{\mathrm{int}}} = \ldots   -\frac{1}{2}\int d^{4}x_3 \,  \int d^{4}x_4 \, \mathcal{L}_{\mathrm{int}}(x_3) \cdot \mathcal{L}_{\mathrm{int}}(x_4) +\ldots 
\, ,
\end{equation}
where the ellipses denote non-quadratic terms in the interaction Lagrangian
\begin{align}
\mathcal{L}_{\textrm{int}}(z_{i}) &= 
\sum_{l=1}^3\bar{q}^{l}(z_i)X_{l,\mu}\gamma ^{\mu _i}q^{l}(z_i)  \, .
\end{align}
Here, we include both gauge fields in $X_{\mu}$ as
\begin{align}
X_{l,\mu}(z)\equiv g_s B^{a}_{\mu}(z)\frac{\lambda_{a}^C}{2}+e\, e_{q_l}\, A_{\mu}(z) \, ,
\end{align}
with $\lambda_{a}^C$ a Gell-Mann color matrix and $B_{\mu}^a(z)$ and $A_{\mu}(z)$ are, respectively, the gluon and photon fields. The field strength contributions are obtained by Taylor expanding the fields and keeping the first gauge invariant non-zero order,
\begin{align}
A_{\mu}(x)&=\frac{1}{2}\, x^{\nu}F_{\nu\mu}=\frac{i}{2}\, F_{\nu\mu}\lim_{q_{4}\rightarrow 0}\partial_{q_4}^{\nu} e^{-iq_4 x} \, ,\\
B_{\mu}^a(x)&=\frac{1}{2}\, x^{\nu}G_{\nu\mu}^a=\frac{i}{2}\, G_{\nu\mu}^a\lim_{q_{4}\rightarrow 0}\partial _{q_4}^{\nu} e^{-iq_4 x} \, .
\end{align}
Including these interaction terms in the Dyson series and closing the quark lines for the leading (one-loop) order contribution, the color trace is factored out and the associated Wilson coefficient contributions to both field strength tensors are  trivially related to the quark loop (i.e.~the one-loop HLbL tensor with a quark running around in the loop) according to
\begin{align}
\Pi^{\mu_1\mu_2\mu_3\mu_4}
& =
\sum_j \frac{ e_{q_j}^{2}}{e^2} \Bigg\langle e^2 e_{q,j}^2  F_{\nu_3'\mu_3'}F_{\nu_4'\mu_4'} +\frac{1}{2 N_{c}} g_s^2 G^{a}_{\nu_3'\mu_3'}G^{a}_{\nu_4'\mu_4'} \Bigg\rangle^{j,\mu_3\mu_4} 
\nonumber \\
& 
\qquad \quad \times
\left(\frac{-1}{8} \lim_{q_3 , q_4 \rightarrow 0}
\, 
\partial_{q_3}^{\nu_3'} \, \partial_{q_4}^{\nu_4'} \, \Pi_{\mathrm{ql,j}}^{\mu_1\mu_2\mu_3 ' \mu_4 ' }\right)\, .
\end{align}
Here the quark loop is given by 
\begin{equation}
\Pi_{\mathrm{ql},j}^{\mu_1\mu_2\mu_3'\mu_4'}=
\left(-i\int \frac{d^{4}q_{4}}{(2\pi)^{4}}\left(\prod_{i=1}^{4}\int d^{4}x_{i}\, e^{-i q_{i} x_{i}}\right)  \langle 0 | T\left(\prod_{j=1}^{4}J_j^{\mu_{j}}(x_{j})\right)|0\rangle \right)_{\mathrm{quark\,  loop}} \, ,
\end{equation}
with all the quark fields in the currents contracted. This topology is diagrammatically depicted in Fig.~\ref{fig:OPEFIELDLO}.
One can then factorise $q_{4}$ from the matrix element (cf.~Eq.~(\ref{eq:factoriseq4})) and take the soft limit to yield the derivative tensor
\begin{align}\label{eq:oped4fieldstrength}
\nonumber
\lim_{q_4\rightarrow 0} \frac{\partial \Pi^{\mu_1\mu_2\mu_3\nu_4}}{\partial q_{4}^{\mu_4}}
& =
\sum_j \frac{ e_{q_j}^{2}}{8 e^2} \, 
\lim_{q_4 \rightarrow 0} 
\left[ 
\partial^{\nu_4}_{q_4}\, \Bigg\langle e^2 e_{q,j}^2  F_{\nu_3'\mu_3'}F_{\nu_4'\mu_4'} +\frac{1}{2N_{c}} g_s^2 G^{a}_{\nu_3'\mu_3'}G^{a}_{\nu_4'\mu_4'} \Bigg\rangle^{j,\mu_3\mu_4} 
\right]
\\ 
& 
\qquad \quad \times 
\lim_{q_3,q_4 \rightarrow 0} \, \partial_{q_3}^{\nu_3'}\, \partial_{q_4}^{\nu_4'} \, \Pi_{\mathrm{ql,j}}^{\mu_1\mu_2\mu_3 ' \mu_4 ' }
\, .
\end{align}
An alternative way of looking at this is to take the diagrams of Fig.~\ref{fig:OPEFIELDLO} and expand in $q_3,q_4$ small with $q_{1,2}=\pm\hat q-(1/2)\left(q_3+q_4\right)$.

\begin{figure}
    \centering
    \includegraphics{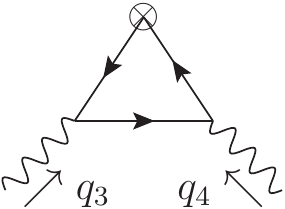}
    \caption{The diagrams needed for the lowest order in $\alpha_s$ matrix elements of the quark operators. The diagrams are with the $q_3,q_4$ legs interchanged must also be taken into account. The crossed circle indicates the insertion of one of the quark operators.}
    \label{fig:matrixLO}
\end{figure}

Also factorising $q_4$ for the quark contributions in Eq.~(\ref{eq:oped4quark}) and adding the result to the field strength piece in Eq.~(\ref{eq:oped4fieldstrength}), we then find at leading order in\footnote{Let us note that while the gluon operator contribution is technically $\alpha_s$-suppressed, and if $Q_3^2\gg \Lambda_{\mathrm{QCD}}^2$ the associated contribution is going to enter suppressed by $\alpha_s^2(Q_{3})$, there is no hard scale flowing through the associated gluon lines, and then the associated contributions do not come with any $\mathcal{O}(\alpha_s(\overline{Q}_3))$ suppression.} $\alpha_{s}(\overline{Q}_3)$ and at NLO in the $Q_3 / \overline{Q}_{3}$ expansion the result
\begin{align}\nonumber
&\lim_{q_4\rightarrow 0} \frac{\partial \Pi^{\mu_1\mu_2\mu_3\nu_4}}{\partial q_{4, \, \mu_4}}
\approx 
\sum_j\frac{ e_{q_ j}^2}{e^2}\,   \lim_{q_4 \rightarrow 0}\partial_{\nu_4}^{q_4}\Bigg\langle\bar{q}(0)\left[
\Gamma^{\mu_1\mu_2}(-\hat{q})-\Gamma^{\mu_2\mu_1}(-\hat{q})
\right]q(0) \Bigg\rangle^{j, \mu_3,\mu_4} 
\\&+\sum_j\frac{ie_{q_j}^2}{e^2\hat{q}^2}\left( g_{\mu_1\delta}g_{\mu_2\beta}+g_{\mu_2\delta}g_{\mu_1\beta}-g_{\mu_1\mu_2}g_{\delta\beta}\right) \left(g^{\alpha\delta}-2\frac{\hat{q}^\delta \hat{q}^\alpha}{\hat{q}^2}\right)
\nonumber \\
& \qquad \qquad \qquad  \times \lim_{q_4 \rightarrow 0}\partial_{\nu_4}^{q_4}\Bigg\langle \bar{q}(0)\left[ 
\overrightarrow{D}^{\alpha}-\overleftarrow{D}^{\alpha}
\right]
\gamma^{\beta}  q(0) \Bigg\rangle ^{j, \mu_3,\, \mu_4} \nonumber \\
&+ \sum _j \frac{ e_{q_j}^{2}}{8 e^2} \lim_{q_4 \rightarrow 0} \left[ \partial^{\nu_4}_{q_4}\Bigg\langle e^2 e_{q,j}^2  F_{\nu_3'\mu_3'}F_{\nu_4'\mu_4'} +\frac{1}{2N_{c}} g_s^2 G^{a}_{\nu_3'\mu_3'}G^{a}_{\nu_4'\mu_4'} \Bigg\rangle^{j,\mu_3\mu_4} \right]
\nonumber \\
& \qquad \qquad \qquad \times
\lim_{q_3,q_4 \rightarrow 0}\partial_{q_3}^{\nu_3'}\, \partial_{q_4}^{\nu_4'}\, 
\Pi_{\mathrm{ql,j}}^{\mu_1\mu_2\mu_3\mu_4}
\, . \label{eq:masterope}
\end{align}
At this stage the dependence on the two independent momenta, $\hat{q}$ and $q_3$, appears completely factorized. The dependence on $\hat q$ is in the Wilson coefficients, the dependence on $q_3$ in the matrix elements of the operators. The matrix elements derived below are pictorially shown as a Feynman diagram in Fig.~\ref{fig:matrixLO}.

\section{Perturbative regime}
\label{sec:alphaSLO}

If $Q_3^2 \gg \Lambda_{\mathrm{QCD}}^2$, the matrix elements of the operators in Eq.~(\ref{eq:masterope}) can be computed perturbatively. In this section we describe the process in some detail, also to illustrate some of the general features of the OPE that can be trivially generalized to the nonperturbative case, such as the need of renormalization and the exact power counting at each step of the calculation, from the Lorentz decomposed HLbL tensor to the final $\hat{\Pi}_{i}$.

\subsection{Dimension $D=3$}
At dimension $D=3$ we have
\begin{equation}
\lim_{q_4\rightarrow 0} \frac{\partial \Pi^{\mu_1\mu_2\mu_3\nu_4}}{\partial q_{4, \, \mu_4}}
=
\sum_j \frac{ e_{q_j}^{2}}{e^2} \lim_{q_4 \rightarrow 0}\partial ^{\nu_4}_{q_4}
\, \Bigg\langle\bar{q}(0)\left[ \Gamma^{\mu_1\mu_2}(-\hat{q})-\Gamma^{\mu_2\mu_1}(-\hat{q})\right] 
q(0) \Bigg\rangle^{j, \mu_3,\mu_4}
\, .
\end{equation}
Expanding two electromagnetic vertices from the action one has
\begin{align}
\label{eq:D3matrix}
& \lim_{q_4\rightarrow 0}\frac{\partial\Pi^{\mu_1\mu_2\mu_3\nu_4}}{\partial q_{4,\, \mu_4}}
 =
\nonumber \\
& =
\sum_{j} N_c \, e_{q,j}^4
\lim_{q_4\rightarrow 0} i \, \partial_{q_4}^{\nu_4}\int \frac{d^d p}{(2\pi)^d}\, \mathrm{Tr}\left[\Gamma^{\mu_1\mu_2}(\hat{q})S(p+q_3+q_4)\gamma^{\mu_3}S(p+q_4)\gamma^{\mu_4}S(p) +(3\leftrightarrow 4)\right] \, ,
\end{align}
We may here take the derivative before integrating and use that for an arbitrary momentum $k$ the propagator satisfies
\begin{equation}
\partial_{q_4}^{\nu_4}S(k+q_{4})=-S(k+q_{4})\gamma^{\nu_4}S(k+q_4) \, .
\end{equation}
In the $q_4 \rightarrow 0$ limit, the remaining one-loop integral depending on only the scale $q_3$ is then straightforward to compute. Alternatively one can evaluate 
\begin{equation}
\lim_{q_4\rightarrow 0}\frac{\partial}{\partial q_{4,\, \mu_4}}
\langle 0
\vert
\bar{q}(0)\left(\gamma^{\mu_1}\gamma^\alpha\gamma^{\mu_2}-\gamma^{\mu_2}\gamma^\alpha\gamma^{\mu_1}\right)q(0)
\vert \gamma^{\mu_3}(q_3)\gamma^{\nu_4}(q_4)\rangle
\, ,
\end{equation}
via the Feynman diagrams of Fig.~\ref{fig:matrixLO} and read off the Wilson coefficient needed from Eq.~(\ref{eq:D3matrix}).

At this level all the terms go as $\mathcal{O}(1/\hat Q)$. A complication arises when extracting the corresponding $\hat{\Pi}_j$ needed for $a_\mu ^{\textrm{HLbL}}$ in the integration in Eq.~(\ref{eq:amuint}). The procedure followed in Ref.~\cite{Bijnens:2021jqo} consisted in first extracting an intermediate set of $19$ independent scalar structures $\tilde{\Pi}_{j}$ in the tensor using corresponding projectors $P^{\tilde{\Pi}_{j}}$. These $\tilde{\Pi}_{j}$ can then be mapped onto the sought $\hat{\Pi}_{j}$ as in Ref.~\cite{Bijnens:2021jqo}. Examples of the projectors to the $\tilde{\Pi}_{j}$ are
$P^{\tilde{\Pi}_{1}}_{\mu_1 \mu_2 \mu_3 \mu_4 \nu_4}=g_{\mu_1\mu_2}\,g_{\mu_3\nu_4}\;q_{1,\mu_4}\,$ and $P^{\tilde{\Pi}_{4}}_{\mu_1 \mu_2 \mu_3 \mu_4 \nu_4}=g_{\mu_1\mu_2}\,g_{\mu_3\nu_4}\;q_{2,\mu_4}$
, with the rest given in Ref.~\cite{Bijnens:2021jqo}. From these projectors, it is clear up to which order one can unambiguously predict each $\tilde{\Pi}_j$: since $q_{1}=\hat{q}-\frac{q_{3}}{2}$ and $q_{2}=-\hat{q}-\frac{q_3}{2}$, it is easy to e.g.~see that the dimension $D=3$ contribution is guaranteed to give the correct answer for $\tilde{\Pi}_{1}$ and $\tilde{\Pi}_{4}$ up to and including order $\left(\frac{1}{\overline{Q}_3}\right)^{0}$.
Obviously, a linear combination of $\tilde{\Pi}_j$ is unambiguously predicted at least up to the most limiting of the addends. However further unambiguous terms can be hidden through cancellation in linear combinations, which become lost if the $\tilde{\Pi}_j$ have already been truncated. A simple example is $\tilde{\Pi}_{1}+\tilde{\Pi}_4$, which is unambiguously predicted up to order $\mathcal{O}\left(\frac{1}{\overline{Q}_3}\right)$.

In order to avoid the problem, we build a new set of $\tilde{\Pi}_{q_3, \, j}$ with $q_3$ and $\hat{q}$ as variables, thus assuring we have a set with as low possible powers of $\hat Q$ in the projectors. The advantage of this basis is that the linear combination of $\tilde{\Pi}_{q_3,\, j}$ is unambiguously predicted at up to the most limiting of the addends, without missing possible kinematic cancellations in the sum (no combination of $\hat{q}$ and $q_3$ can lower the naive $\mathcal{O}\left(\frac{1}{\overline{Q}_3}\right)$ order). The new projectors to this basis are
\begin{align}
P^{\tilde{\Pi}_{1}}_{q_3} &= g_{\mu_1\mu_2} \, g_ {\mu_3\nu_4} \, \hat{q}_{\mu_4}\,  \\  
P^{\tilde{\Pi}_{2}}_{q_3} &= g_{\mu_1\mu_2} \, g_ {\mu_3\nu_4} q_{3\mu_4}\, , \\  
P^{\tilde{\Pi}_{3}}_{q_3} &= g_{\mu_1\mu_3} \, g_ {\mu_2\nu_4} \, \hat{q}_{\mu_4}\, , \\  
P^{\tilde{\Pi}_{4}}_{q_3} &= g_{\mu_1\mu_3} \, g_ {\mu_2\nu_4} q_{3\mu_4}\, , \\  
P^{\tilde{\Pi}_{5}}_{q_3} &= g_{\mu_2\mu_3} \, g_ {\mu_1\nu_4} \, \hat{q}_{\mu_4}\, , \\  
P^{\tilde{\Pi}_{6}}_{q_3} &= g_{\mu_2\mu_3} \, g_ {\mu_1\nu_4} q_{3\mu_4}\, , \\  
P^{\tilde{\Pi}_{7}}_{q_3} &= g_{\mu_1\nu_4} \, g_ {\mu_2\mu_4} \, \hat{q}_{\mu_3}\, , \\  
P^{\tilde{\Pi}_{8}}_{q_3} &= g_{\mu_2\nu_4} \, g_ {\mu_3\mu_4} q_{3\mu_1}\, , \\  
P^{\tilde{\Pi}_{9}}_{q_3} &= g_{\mu_3\nu_4} \, g_ {\mu_1\mu_4} q_{3\mu_2}\, , \\  
P^{\tilde{\Pi}_{10}}_{q_3} &=  g_{\mu_1\mu_2} \, \hat{q}_{\mu_3} \, \hat{q}_{\nu_4} q_{3\mu_4}\, , \\  
P^{\tilde{\Pi}_{11}}_{q_3} &= g_{\mu_2\mu_3} q_{3\mu_1} \, \hat{q}_{\nu_4} q_{3\mu_4}\, , \\  
P^{\tilde{\Pi}_{12}}_{q_3} &= g_{\mu_3\mu_1} q_{3\mu_2} q_{3 \nu_4} \, \hat{q}_{\mu_4}\, , \\  
P^{\tilde{\Pi}_{13}}_{q_3} &= g_{\mu_1\nu_4} q_{3\mu_2} \, \hat{q}_{\mu_3} \, \hat{q}_{\mu_4}\, , \\  
P^{\tilde{\Pi}_{14}}_{q_3} &= g_{\mu_1\nu_4} q_{3\mu_2} \, \hat{q}_{\mu_3} q_{3\mu_4}\, , \\  
P^{\tilde{\Pi}_{15}}_{q_3} &= g_{\mu_2\nu_4} q_{3\mu_1} \, \hat{q}_{\mu_3} \, \hat{q}_{\mu_4}\, , \\  
P^{\tilde{\Pi}_{16}}_{q_3} &= g_{\mu_2\nu_4} q_{3\mu_1} \, \hat{q}_{\mu_3} q_{3\mu_4}\, , \\  
P^{\tilde{\Pi}_{17}}_{q_3} &= g_{\mu_3\nu_4} q_{3\mu_1} q_{3\mu_2} \, \hat{q}_{\mu_4}\, , \\  
P^{\tilde{\Pi}_{18}}_{q_3} &= g_{\mu_3\nu_4} q_{3\mu_1} q_{3\mu_2} q_{3\mu_4}\, , \\  
P^{\tilde{\Pi}_{19}}_{q_3} &= q_{3\mu_1} q_{3\mu_2} \, \hat{q}_{\mu_3} \, \hat{q}_{\nu_4} q_{3\mu_4} \, ,
\end{align}
from which relations can be found between the old $ \tilde{\Pi}_j$ and the new $ \tilde{\Pi}_{q_3 , \, j}$, as needed to reach the $\hat{\Pi}_{j}$. We repeat exactly the same procedure for the case in which $q_{1}$ or $q_2$ is the small momentum. Notice how building the new $q_1$ or $q_2$ projectors can be done by cyclic permutation, i.e.
by changing $q_3\rightarrow q_1$ and $\hat q= \left(q_1-q_2\right) /2 \rightarrow \hat q_{23}=\left(q_2-q_3\right) /2$ for the $q_1$ small case, and 
 $q_3\rightarrow q_2$ and $\hat q=\left(q_1-q_2\right) /2\rightarrow \hat q_{31}= \left(q_3-q_1\right) / 2$ for the $q_2$ small case.

At this stage it is straightforward to extract the linear combinations of $\tilde{\Pi}_{q_3, \, j}$ that can enter order by order in the $\hat{\Pi}_{j}$. We find that truncating at order $D$ in the operators one can unambiguously predict,
\begin{align}
& 
\hat{\Pi}_1 \rightarrow \frac{1}{\overline{Q}_3^{D-1}} \, , \quad 
\hat{\Pi}_4 \rightarrow \frac{1}{\overline{Q}_3^{D}} \, , \quad
\hat{\Pi}_7 \rightarrow \frac{1}{\overline{Q}_3^{D+1}}  \, , \quad
\hat{\Pi}_{17} \rightarrow \frac{1}{\overline{Q}_3^{D}}  \, , \nonumber \\
&
\hat{\Pi}_{39} \rightarrow \frac{1}{\overline{Q}_3^{D}} \, , \quad \hat{\Pi}_{54} \rightarrow \frac{1}{\overline{Q}_3^{D}} \, .
\end{align}
At $D=3$ one then finds for $q_3$ small
\begin{align}\label{eq:d3quarkq3}
\underline{q_{3}\textrm{ small:}} &
\nonumber \\
\hat{\Pi}_1&=-\frac{4N_{c}\sum_{j}e_{q,j}^4}{\pi^2 Q_{3}^2\, \overline{Q}_3^2} +\mathcal{O} \left(\frac{1}{\overline{Q}_3^{3}}\right) \, , \, \hat{\Pi}_4= \mathcal{O} \left(\frac{1}{\overline{Q}_3^{4}}\right) \, , \, \hat{\Pi}_7= \mathcal{O} \left(\frac{1}{\overline{Q}_3^{5}} \right) \, , \,\\ \hat{\Pi}_{17}&= \mathcal{O} \left(\frac{1}{\overline{Q}_3^{4}}\right)  \, , \, \hat{\Pi}_{39}= \mathcal{O} \left(\frac{1}{\overline{Q}_3^{4}}\right) \, , \, \hat{\Pi}_{54}= \mathcal{O} \left(\frac{1}{\overline{Q}_3^{4}} \right)\, .
\end{align}
For $q_1$ small the result is
\begin{align}\label{eq:d3quarkq1}
\underline{q_{1}\textrm{ small:}} &
\nonumber \\
\hat{\Pi}_1&=\mathcal{O} \left(\frac{1}{\overline{Q}_1^{4}}\right) \, , \, \hat{\Pi}_4= \mathcal{O} \left(\frac{1}{\overline{Q}_1^{2}}\right) \, , \, \hat{\Pi}_7= \mathcal{O} \left(\frac{1}{\overline{Q}_1^{3}} \right) \, , \,\\ \hat{\Pi}_{17}&= \mathcal{O} \left(\frac{1}{\overline{Q}_1^{4}}\right)  \, , \, \hat{\Pi}_{39}= \mathcal{O} \left(\frac{1}{\overline{Q}_1^{4}}\right) \, , \, \hat{\Pi}_{54}= \mathcal{O} \left(\frac{1}{\overline{Q}_1^{4}} \right)\, .
\end{align}
Finally, for $q_2$ small we find
\begin{align}\label{eq:d3quarkq2}
\underline{q_{2}\textrm{ small:}} &
\nonumber \\
\hat{\Pi}_1&=\mathcal{O} \left(\frac{1}{\overline{Q}_2^{4}}\right) \, , \, \hat{\Pi}_4= \mathcal{O} \left(\frac{1}{\overline{Q}_2^{2}}\right) \, , \, \hat{\Pi}_7= \mathcal{O} \left(\frac{1}{\overline{Q}_2^{4}} \right) \, , \,\\ \hat{\Pi}_{17}&= \mathcal{O} \left(\frac{1}{\overline{Q}_2^{4}}\right)  \, , \, \hat{\Pi}_{39}= \mathcal{O} \left(\frac{1}{\overline{Q}_2^{4}}\right) \, , \, \hat{\Pi}_{54}= \mathcal{O} \left(\frac{1}{\overline{Q}_2^{4}} \right)\, .
\end{align}

\subsection{Dimension $D=4$}
The first operator appearing at dimension $D=4$ in Eq.~(\ref{eq:masterope}) involves a quark current. Expanding down two electromagnetic vertices from the Dyson series in the matrix element and contracting one finds, in the perturbative limit $Q_3^2 \gg \Lambda _{\textrm{QCD}}^2 $,
\begin{align}
&\lim_{q_4 \rightarrow 0}\partial ^{\nu_4}_{q_4} \, \Bigg\langle \bar{q}(0)
\left[ 
\overrightarrow{D}^{\alpha}-\overleftarrow{D}^{\alpha}
\right] 
\gamma^{\beta}  q(0) \Bigg\rangle ^{j, \mu_3,\, \mu_4}  
\nonumber \\ & 
=-2 N_c\, e^2 e_{q,j}^2\, \lim_{q_4 \rightarrow 0}\, \partial ^{\nu_4}_{q_4}\int \frac{d^{4}k}{(2\pi)^4}\, k^{\alpha}\, \mathrm{Tr}[\gamma^{\beta}S(k+q_3+q_4)\gamma^{\mu_3}S(k+q_4)\gamma^{\mu_4}S(k)+(3 \leftrightarrow 4) ]\, . \label{eq:MEderpert}
\end{align}
The result for the matrix element is divergent at $d=4$. We regularize the divergence within dimensional regularization, with $d=4-2\epsilon$. 
Inserting Eq.~\eqref{eq:MEderpert} into Eq.~\eqref{eq:masterope}, one finds the derivative of the HLbL tensor, where the divergence in the matrix element in Eq.~(\ref{eq:MEderpert}) cancels the divergence from the Wilson coefficient of the field strength contribution at dimension $D=4$ in Eq.~(\ref{eq:masterope}), thus yielding a finite result. 

An alternative approach is to separately project the quark operator matrix element and the photon matrix element on the $\hat\Pi_i$. As we will explain below, one also has to re-expand the $D=3$ contribution in Eq.~(\ref{eq:masterope}). In the following we study the three OPE contributions separately.

\subsubsection{Dimension $D=4$ quark current operator}
The $\hat{\Pi}_i$ are finite in the $q_3$-small regime. Leaving out the overall factor $N_c \sum _j e_{q_j}^4$ for the $\hat{\Pi}_i$ here and in the following, one finds
\begin{align}\label{eq:d4quarkq3}
\underline{q_{3}\textrm{ small:}} &
\nonumber \\
\nonumber
\hat{\Pi}_1
&=
\mathcal{O} \left(\frac{1}{\overline{Q}_3^{4}}\right) 
\, , \qquad \qquad \qquad 
\hat{\Pi}_{4\, }
=
\frac{16(\delta_{12}^4-\delta_{12}^2 Q_{3}^2+Q_{3}^4)}{3\pi^2(Q_{3}^2-\delta_{12}^2)^2\overline{Q}_3^{4}} 
\, , \, 
\nonumber \\
\hat{\Pi}_7
& =
\frac{32\delta_{12}Q_{3}^2}{3\pi^2(Q_{3}^2-\delta_{12}^2)^2 \overline{Q}_{3}^5}
\, , \, \qquad 
\hat{\Pi}_{17}
=
\frac{4(Q_{3}^2-4\delta_{12}^2)}{3\pi^2 Q_{3}^2(Q_{3}^2-\delta_{12}^2)\overline{Q}_{3}^4} 
\, , \, 
\nonumber \\
\hat{\Pi}_{39}
& = 
\frac{4(Q_{3}^4+4\delta_{12}^4-9\delta_{12}^2 Q_{3}^2)}{3\pi^2 Q_{3}^2(Q_{3}^2-\delta_{12}^2)^2\overline{Q}_{3}^4}
\, , \, 
\hat{\Pi}_{54}
=
\mathcal{O} \left(\frac{1}{\overline{Q}_3^{5}} \right)\, .
\end{align}
It is important to note that there are kinematic singularities in the limit $\delta_{12}\rightarrow Q_{3}$. Since the $\hat{\Pi}_i$ by construction should be free of such singularities, they must cancel against the remaining contribution from the OPE in Eq.~(\ref{eq:masterope}), thus connecting the different $D=3$ and $D=4$ Green functions. In this section we will show how this explicitly occurs in the perturbative regime. The fact that it must also occur in the nonperturbative regime is going to lead to simple nontrivial relations among the different form factors appearing in Lorentz decompositions of the nonperturbative matrix elements, as explained in Sec.~\ref{sec:nonperturbative}.

An additional feature for the $\hat{\Pi}_i$ appears when $q_{1}$ or $q_{2}$ is small. With $1/\hat{\epsilon} = 1/\epsilon -\gamma _{\textrm{E}}+\log 4\pi$ and neglecting terms of order $\epsilon$ and higher, one finds
\begin{align}\label{eq:d4quarkq1}
\nonumber
\underline{q_{1}\textrm{ small:}} &
\nonumber \\
\hat{\Pi}_1
&
=-\frac{16[3+ \epsilon (-4-6\ln Q_{1})]}{9\pi^2\hat{\epsilon}\, \overline{Q}_{1}^4} 
\, , \, 
\hat{\Pi}_4
=
-\frac{4\delta_{23}^4-9\delta_{23}^2Q_{1}^2+Q_{1}^4}{3\pi^2 Q_{1}^2(Q_{1}^2-\delta_{23}^2)^2\overline{Q}_{1}^2} 
\, , \,
\nonumber \\ 
\hat{\Pi}_7
& =
4 \frac{3\delta_{23}^3-5\delta_{23}Q_{1}^2}{3\pi^2 Q_{1}^2(Q_{1}^2-\delta_{23}^2)^2\overline{Q}_{1}^3} 
\, , \,
\quad 
\hat{\Pi}_{17}
=
\mathcal{O} \left(\frac{1}{\overline{Q}_1^{5}}\right)  \, , \,
\nonumber \\
 \hat{\Pi}_{39}
 & =
 4\frac{4\delta_{23}^4-9\delta_{23}^2Q_{1}^2+Q_{1}^4}{3\pi^2 Q_{1}^2(Q_{1}^2-\delta_{23}^2)^2\overline{Q}_{1}^4} 
 \, , \, 
 \quad
 \hat{\Pi}_{54}
 =
 -4\frac{Q_{1}^2+2\delta_{23}^2}{3\pi^2 Q_{1}^2(Q_{1}^2-\delta_{23}^2)\overline{Q}_{1}^4}  
 \, .
\end{align}
\begin{align}\label{eq:d4quarkq2}
\underline{q_{2}\textrm{ small:}} &
\nonumber \\
\nonumber
\hat{\Pi}_1
&
=
-\frac{16[3+\epsilon (-4-6\ln Q_{2})]}{9\pi^2\hat{\epsilon}\, \overline{Q}_2^4} 
\, , \, 
\hat{\Pi}_4
=
-\frac{4\delta_{31}^4-9\delta_{31}^2Q_2^2+Q_2^4}{3\pi^2 Q_2^2(Q_2^2-\delta_{31}^2)^2\overline{Q}_2^2}  
\, , \,
\nonumber \\
\hat{\Pi}_7
& 
=
-4 \frac{4\delta_{31}^4-9\delta_{31}^2 Q_{2}^2+Q_{2}^4}{3\pi^2 Q_2^2(Q_2^2-\delta_{31}^2)^2\overline{Q}_2^4} \, , \,
\hat{\Pi}_{17}
=
\mathcal{O} \left(\frac{1}{\overline{Q}_2^{5}}\right)  \, , \,
\nonumber \\
\hat{\Pi}_{39}
&
=4\frac{4\delta_{31}^4-9\delta_{31}^2Q_2^2+Q_2^4}{3\pi^2 Q_2^2(Q_2^2-\delta_{31}^2)^2\overline{Q}_2^4} 
\, , \, 
\quad 
\hat{\Pi}_{54}
=
4\frac{Q_2^2+2\delta_{31}^2}{3\pi^2 Q_2^2(Q_2^2-\delta_{31}^2)\overline{Q}_2^4}  \, .
\end{align}
As can be seen, $\hat{\Pi}_1$ contains a UV divergent logarithm in both limits.

\subsubsection{Dimension $D=4$ photon field strength operator}
Taking into account that at the studied order no other matrix elements are divergent, the only way of cancelling the logarithmic divergences in Eqs.~(\ref{eq:d4quarkq3})--(\ref{eq:d4quarkq2}) is with a similar (in this case infrared) logarithm in a Wilson coefficient. This comes from the photon field strength contribution, 
\begin{align}\nonumber
&\lim_{q_4\rightarrow 0} \frac{\partial \Pi^{\mu_1\mu_2\mu_3\nu_4}}{\partial q_{4, \, \mu_4}}= \frac{\sum_j e_{q_j}^4}{8} \, \lim_{q_4 \rightarrow 0}
\left[
\partial^{\nu_4}_{q_4}
\, \left\langle  F_{\nu_3'\mu_3'}F_{\nu_4'\mu_4'} \right\rangle^{\mu_3\mu_4}
\right] \, 
\lim_{q_3,q_4 \rightarrow 0}\partial_{q_3}^{\nu_3'}\, \partial_{q_4}^{\nu_4'}\, \Pi_{\mathrm{ql}}^{\mu_1\mu_2\mu_3 ' \mu_4 ' }
\, . 
\end{align}
Since photons do not see strong interactions at the studied order in $\alpha$, the computation of the associated matrix element is straightforward. The result is
\begin{align}\nonumber
    \lim_{q_4 \rightarrow 0}\partial^{\nu_4}_{q_4}\, \left\langle  F_{\nu_3'\mu_3'}F_{\nu_4'\mu_4'} \right\rangle^{\mu_3\mu_4}=&-\Bigg[
\left( q_{3,\, \mu_3'}\, g_{\nu_3'}^{ \mu_3}-q_{3,\, \nu_3'}\, g_{\mu_3'}^{\mu_3}\right)
\left(g_{\mu_4'}^{\nu_4}\, g_{\nu_4'}^{\mu_4}-g_{\nu_4'}^{ \nu_4}\, g_{\mu_4'}^{\mu_4}\right)
\\
&
\quad 
+\left( g_{\mu_3'}^{\nu_4}\, g_{\nu_3'}^{\mu_4}-g_{\nu_3'}^{\nu_4}\, g_{\mu_3'}^{\mu_4}\right)
\left( q_{3,\, \mu_4'}\, g_{\nu_4'}^{\mu_3}-q_{3,\, \nu_4'}\, g_{\mu_4'}^{\mu_3}\right)
\Bigg]  \, .
\end{align}
The difficulty here lies in the computation of
\begin{align}
    \lim_{q_3,q_4 \rightarrow 0} \partial_{q_3}^{\nu_3'} \, \partial_{q_4}^{\nu_4'}\, \Pi_{\mathrm{ql}}^{\mu_1\mu_2\mu_3 ' \mu_4' } \, .
\end{align}
The massless quark loop tensor, $\Pi_{\mathrm{ql}}^{\mu_1\mu_2\mu_3' \mu_4' }$ is well-known to be finite and also the derivative $\lim_{q_4 \rightarrow 0}\partial_{q_4}^{\nu_4'}\, \Pi_{\mathrm{ql}}^{\mu_1\mu_2\mu_3\mu_4'}$. However, if one tries to apply the remaining $\lim_{q_3 \rightarrow 0}\partial_{q_3}^{\nu_3'}$ operation to the known analytic expression of $\lim_{q_4 \rightarrow 0}\partial_{q_4}^{\nu_4'}\Pi_{\mathrm{ql}}^{\mu_1\mu_2\mu_3'\mu_4'}$, one finds divergent logarithms. Indeed, these IR divergences at the Wilson coefficient level are the ones that must cancel the corresponding UV divergences of the quark current matrix element discussed above. To recover the correct finite terms, however, one needs to regularize both the Wilson coefficient and the matrix element in the same way, which we do using dimensional regularization. Instead of first performing the multi-scale loop integral and then applying the derivatives and limits $\lim_{q_i \rightarrow 0}\partial_{q_i}^{\nu_i'}$, we once again first apply the operations and then straightforwardly calculate the resulting single-scale massless one-loop integrals. Combining them together one finds the associated derivative HLbL tensor. Projecting them into the $\hat{\Pi}_i$ one respectively obtains in the three regions of small $q_i$
\begin{align}\label{eq:d4photonq3}
\underline{q_{3}\textrm{ small:}} &
\nonumber \\
\nonumber
\hat{\Pi}_1
&
=\mathcal{O} \left(\frac{1}{\overline{Q}_3^{4}}\right) 
\, , \, 
\qquad \qquad \quad
\hat{\Pi}_4
=
-\frac{16(2\delta_{12}^4-3\delta_{12}^2 Q_{3}^2+2Q_{3}^4)}{3\pi^2(Q_{3}^2-\delta_{12}^2)^2\overline{Q}_3^{4}} 
\, , \, 
\nonumber \\
\hat{\Pi}_7
&
=
-\frac{32\delta_{12}Q_{3}^2}{3\pi^2(Q_{3}^2-\delta_{12}^2)^2 \overline{Q}_{3}^5} 
\, , \,
\hat{\Pi}_{17}
= 
\mathcal{O} \left(\frac{1}{\overline{Q}_3^{5}}\right)  
\, , \, 
\nonumber \\
\hat{\Pi}_{39}
& 
= 
\frac{16\delta_{12}^2}{3\pi^2 (Q_{3}^2-\delta_{12}^2)^2\overline{Q}_{3}^4}
\, , \, 
\quad
\hat{\Pi}_{54}
= 
\mathcal{O} \left(\frac{1}{\overline{Q}_3^{5}} \right)\, ,
\end{align}
\begin{align}\label{eq:d4photonq1}
\underline{q_{1}\textrm{ small:}} &
\nonumber \\
\nonumber
\hat{\Pi}_1
&
=
\frac{16[1+\epsilon (-3-2\ln \frac{\overline{Q}_{1}}{2})]}{3\pi^2\hat{\epsilon}\, \overline{Q}_{1}^4} 
\, , \, 
\quad
\hat{\Pi}_4
=
-\frac{4\delta_{23}^2}{3\pi^2 (Q_{1}^2-\delta_{23}^2)^2\overline{Q}_{1}^2} 
\, , \, 
\nonumber \\
\hat{\Pi}_7
&
=
8 \frac{\delta_{23}}{3\pi^2 (Q_{1}^2-\delta_{23}^2)^2\overline{Q}_{1}^3}
\, , \, 
 \qquad 
\hat{\Pi}_{17}
= 
\mathcal{O} \left(\frac{1}{\overline{Q}_1^{5}}\right)  
\, , \, 
\nonumber \\
\hat{\Pi}_{39}
& 
=
\frac{16\,\delta_{23}^2}{3\pi^2 (Q_{1}^2-\delta_{23}^2)^2\overline{Q}_{1}^4} 
\, , \, 
\qquad \quad
\hat{\Pi}_{54}
=
\mathcal{O}\left(\frac{1}{\overline{Q}_1^5} \right) \, ,
\end{align}
\begin{align}\label{eq:d4photonq2}
\underline{q_{2}\textrm{ small:}} &
\nonumber \\
\nonumber
\hat{\Pi}_1
&
=
16\frac{1+\epsilon (-3-2\ln \frac{\overline{Q}_{2}}{2})}{3\pi^2\hat{\epsilon}\, \overline{Q}_2^4}
\, , \, 
\quad 
\hat{\Pi}_4
=
-\frac{4\delta_{31}^2}{3\pi^2 (Q_2^2-\delta_{31}^2)^2\overline{Q}_2^2}
\, , \, 
\nonumber \\
\hat{\Pi}_7
&
=
- \frac{16\,\delta_{31}^2}{3\pi^2 (Q_2^2-\delta_{31}^2)^2\overline{Q}_2^2}
\, , \,
\quad 
\hat{\Pi}_{17}
=
\mathcal{O} \left(\frac{1}{\overline{Q}_2^{5}}\right)  
\, , \, 
\nonumber \\
\hat{\Pi}_{39}
 & =
\frac{16\,\delta_{31}^2}{3\pi^2 (Q_2^2-\delta_{31}^2)^2\overline{Q}_2^4}
\, , \, 
\quad \quad
\hat{\Pi}_{54}
=
\mathcal{O} \left(\frac{1}{\overline{Q}_2^{5}}\right)  \, .
\end{align}
Comparing to the $\hat{\Pi}_i$ for the quark-current matrix element in Eqs.~(\ref{eq:d4quarkq3})--(\ref{eq:d4quarkq2}), one notices that the divergence in $\epsilon$ cancels when adding the contributions fro the quark and field strength operators. The infrared and ultraviolet logarithms nicely recombine into $\log\frac{\overline{Q}_i}{2Q_i}$. However, the collinear divergences when $\delta _{ij}\rightarrow Q_k$ remain. 

\subsubsection{Expanded dimension $D=3$ operator}
To cancel the collinear divergences, one also has to add the previously computed $D=3$ expanded to NLO in the energy expansion for the $\hat{\Pi}_i $. The $\hat{\Pi}_i$ in the three momentum regimes, including also the previous LO $\hat{\Pi}_1$ contribution from Eq.~(\ref{eq:d3quarkq3}), are given by
\begin{align}\label{eq:d3expquarkq3}
\underline{q_{3}\textrm{ small:}} &
\nonumber \\
\nonumber
\hat{\Pi}_1&=-\frac{4}{\pi^2\, Q_{3}^2\,\overline{Q}_{3}^2}
\, , \, \quad
\hat{\Pi}_4= \mathcal{O} \left(\frac{1}{\overline{Q}_3^{5}}\right)
\, , \, \quad
\hat{\Pi}_7= \mathcal{O} \left(\frac{1}{\overline{Q}_3^{6}} \right) 
\, , \,
\\ 
\hat{\Pi}_{17}&=\frac{4}{\pi^2(Q_{3}^2-\delta_{12}^2)\overline{Q}_{3}^4}
\, , \, \quad
\hat{\Pi}_{39}= \frac{4}{\pi^2(Q_{3}^2-\delta_{12}^2)\overline{Q}_{3}^4} 
\, , \, \quad
\hat{\Pi}_{54}= \mathcal{O} \left(\frac{1}{\overline{Q}_3^{5}} \right)\, .
\end{align}
\begin{align}\label{eq:d3expquarkq1}
\underline{q_{1}\textrm{ small:}} &
\nonumber \\
\nonumber
\hat{\Pi}_1&=\mathcal{O}\left(\frac{1}{\overline{Q}_{1}^5} \right) 
\, , \, \quad
\hat{\Pi}_4=-\frac{1}{\pi^2 (Q_{1}^2-\delta_{23}^2)\overline{Q}_{1}^2} 
\, , \, \quad
\hat{\Pi}_7=\frac{4\delta_{23}}{\pi^2 (Q_{1}^2-\delta_{23}^2) Q_{1}^2\overline{Q}_{1}^3} 
\, , \,
\\ 
\hat{\Pi}_{17}&= \mathcal{O}\left(\frac{1}{\overline{Q}_3^5} \right)  
\, , \, \quad
\hat{\Pi}_{39}=\frac{4}{\pi^2 (Q_{1}^2-\delta_{23})\overline{Q}_{1}^4} 
\, , \, \quad
\hat{\Pi}_{54}=\frac{4}{\pi^2 (Q_{1}^2-\delta_{23}^2)\overline{Q}_{1}^4} \, .
\end{align}
\begin{align}\label{eq:d3expquarkq2}
\underline{q_{2}\textrm{ small:}} &
\nonumber \\
\nonumber
\hat{\Pi}_1&=\mathcal{O}\left(\frac{1}{\overline{Q}_{2}^5} \right)
\, ,\, \quad
\hat{\Pi}_{17}= \mathcal{O}\left(\frac{1}{\overline{Q}_{2}^5} \right)
\, , \, \quad 
\hat{\Pi}_4=-\frac{1}{\pi^2 (Q_2^2-\delta_{31}^2)\overline{Q}_2^2} 
\, ,\, 
\nonumber \\ 
\hat{\Pi}_7 & =-\hat{\Pi}_{39}=\hat{\Pi}_{54}=-\frac{4}{\pi^2 (Q_2^2-\delta_{31}^2)\overline{Q}_2^4}  \, .
\end{align}

\subsubsection{Final $\hat{\Pi}_i$ at dimension $D=4$}
At this stage we can add the three contributions together (Eqs.~(\ref{eq:d4quarkq3}--\ref{eq:d4photonq2}) for the quark current operator, Eqs.~(\ref{eq:d4photonq3}--\ref{eq:d4photonq2}) for the photon operator and Eqs.~(\ref{eq:d3expquarkq3}--\ref{eq:d3expquarkq2})) in the three momentum limits to get a finite result at NLO in the $\frac{1}{\overline{Q}_3}$ expansion. The results are
\begin{align}\label{eq:d4pihatfinalq3}
\underline{q_{3}\textrm{ small:}} &
\nonumber \\
\nonumber
\hat{\Pi}_1&=-\frac{4}{\pi^2\, Q_{3}^2\,\overline{Q}_{3}^2}
\quad
\hat{\Pi}_4= -\frac{16}{3\pi^2 \overline{Q}_{3}^4}
\, , \, 
\quad
\hat{\Pi}_7= \mathcal{O} \left(\frac{1}{\overline{Q}_3^{6}} \right) \, , \,\\ 
\hat{\Pi}_{17}&=\frac{16}{3\pi^2 Q_{3}^2\overline{Q}_{3}^4}
\, , \, 
\quad
\hat{\Pi}_{39}= \frac{16}{3 \pi^2 Q_{3}^2\overline{Q}_{3}^4} \, , \, 
\quad
\hat{\Pi}_{54}= \mathcal{O} \left(\frac{1}{\overline{Q}_3^{5}} \right)\, ,
\end{align}
\begin{align}\label{eq:d4pihatfinalq1}
\underline{q_{1}\textrm{ small:}} &
\nonumber \\
\nonumber
\hat{\Pi}_1&=-\frac{16\left(5+6\ln\frac{\overline{Q}_1}{2Q_1}\right)}{9\pi^2\overline{Q}_{1}^4}
\, , \, 
\quad
\hat{\Pi}_4=-\frac{4}{3\pi^2 Q_{1}^2\overline{Q}_{1}^2} 
\, , \, 
\nonumber \\
\hat{\Pi}_7 & =\mathcal{O}\left(\frac{1}{\overline{Q}_{1}^4} \right) \, , \,
\quad \qquad
\hat{\Pi}_{17}=\mathcal{O}\left(\frac{1}{\overline{Q}_{1}^5} \right) \, , \, 
\nonumber \\
\hat{\Pi}_{39} & =\frac{16}{3\pi^2 Q_{1}^2\overline{Q}_{1}^4} 
\, , \, 
\quad
\hat{\Pi}_{54}=\frac{8}{3\pi^2 Q_{1}^2\overline{Q}_{1}^4} 
\, ,
\end{align}
\begin{align}\label{eq:d4pihatfinalq2}
\underline{q_{2}\textrm{ small:}} &
\nonumber \\
\nonumber
\hat{\Pi}_1&=-\frac{16\left(5+6\ln \frac{\overline{Q}_{2}}{2Q_2}\right)}{9\pi^2 \overline{Q}_2^4}
\, , \,
\quad
\hat{\Pi}_4=-\frac{4}{3\pi^2 Q_2^2\overline{Q}_2^2}  \, , \, 
\nonumber \\
\hat{\Pi}_7 & =- \frac{16}{3\pi^2 Q_2^2\overline{Q}_2^4} 
\, , \,
\qquad \qquad
\hat{\Pi}_{17}= \mathcal{O} \left(\frac{1}{\overline{Q}_2^{5}}\right)  \, , \, 
\nonumber \\
\hat{\Pi}_{39}& =\frac{16}{3\pi^2 Q_2^2\overline{Q}_2^4}
\, , \, 
\qquad \qquad \quad 
\hat{\Pi}_{54}=-\frac{8}{3\pi^2 Q_2^2\overline{Q}_2^4}
\, .
\end{align}
The OPE developed in Refs.~\cite{Bijnens:2019ghy,Bijnens:2020xnl,Bijnens:2021jqo} was valid in the purely short-distance limit $Q_{i}^2,Q_{j}^2,Q_{k}^2\gg \Lambda_{\mathrm{QCD}}^2$. Here we have considered the Melnikov-Vainshtein limit $Q_{i}^2,Q_{j}^2\gg Q_{k}^2\gg \Lambda _{\textrm{QCD}}^2$ where also the momentum $Q_k^2$ is in the perturbative regime. This is part of the region\footnote{Notice however how in the very extreme (and phenomenologically irrelevant) case in which $Q_{1,2}^2\gg Q_{3}^2\gg \Lambda_{\mathrm{QCD}}^2$, one may need to resum $\ln\frac{\overline{Q}_3}{2Q_3}$ factors.
}
$Q_{i}^2,Q_{j}^2,Q_{k}^2\gg \Lambda_{\mathrm{QCD}}^2$ and was called the corner region in Ref.~\cite{Bijnens:2021jqo}. The $\hat{\Pi}_i$ in Eqs.~(\ref{eq:d4pihatfinalq3}--\ref{eq:d4pihatfinalq2}) should therefore agree with the corner region $\hat{\Pi}_i$ in Ref.~\cite{Bijnens:2021jqo}. Indeed, all the $\hat{\Pi}_i$ exactly agree.

\subsubsection{Renormalization of the OPE}
Up to now we have worked with divergent matrix elements and Wilson coefficients. One can instead work with both finite matrix elements and Wilson coefficients by performing standard renormalization of the OPE, see e.g.~Ref.~\cite{Buchalla:1995vs}. At first sight it may seem striking that already at the leading order contribution in $\alpha_s$ and $\alpha$ one has to perform such a task for the dimension $D=4$ contribution. This can be regarded as a consequence of an intermediate mismatch in the order in $\alpha$ of the two contributing operators, of the respective forms $ \bar{q}D\Gamma q$ and $\alpha \, F \cdot F$ where $\Gamma$ is a Dirac matrix, which eventually enter at the same order in $\alpha$, since the former needs to expand two extra EM vertices in order to capture the two initial photons. In that intermediate step, the QED mixing between $ \bar{q}D\Gamma q$ and $ \alpha \, F \cdot F$ needs to be considered, even when the self-renormalization of each separate operator is not going to affect our calculation at the studied order.

In order to assess the mixing, one can simply focus on the loop divergences of the relevant matrix element. For the quark current operator we have
\begin{align}\nonumber
&\lim_{q_4 \rightarrow 0}\partial ^{\nu_3}_{q_3}\, \partial ^{\nu_4}_{q_4}\, \Bigg\langle \bar{q}(0)\left[ \overrightarrow{D}^{\alpha}-\overleftarrow{D}^{\alpha}\right] \gamma^{\beta}  q(0) \Bigg\rangle ^{\mu_3,\, \mu_4, j} = {\textrm{finite}} -   i\, \frac{8}{3}\, \frac{\alpha}{4\pi}\, N_c\, e_{q,j}^2 \, \frac{ Q_{3}^{-2\epsilon}}{ \epsilon}
\\
& 
\times \nonumber
\Big(
- g^{\mu_3\mu_4}g^{\alpha\beta}g^{\nu_3\nu_4} 
+ g^{\mu_3\mu_4}   g^{\alpha\nu_3}g^{\beta\nu_4} 
+ g^{\mu_3\mu_4}g^{\alpha\nu_4}g^{\beta\nu_3} 
+    g^{\mu_3\alpha}g^{\mu_4\beta}g^{\nu_3\nu_4} 
- g^{\mu_3\alpha}g^{\mu_4\nu_3}g^{\beta\nu_4}    
\\
&
+ g^{\mu_3\beta}g^{\mu_4\alpha}g^{\nu_3\nu_4} 
- g^{\mu_3\beta}g^{\mu_4\nu_3}   g^{\alpha\nu_4}
- g^{\mu_3\nu_4}g^{\mu_4\alpha}g^{\beta\nu_3} 
- g^{\mu_3\nu_4}   g^{\mu_4\beta}g^{\alpha\nu_3} 
+ g^{\mu_3\nu_4}g^{\mu_4\nu_3}g^{\alpha\beta}
\Big)
 \nonumber
\\
&
=
\textrm{finite} - i\, \frac{2}{3}\,  N_c\, e_{q,j}^2 \, \frac{ Q_{3}^{-2\epsilon}}{\hat{\epsilon}}\lim_{q_4 \rightarrow 0}\, \partial^{\nu_3}_{q_3}\, \partial^{\nu_4}_{q_4}\, \Bigg\langle  \frac{\alpha}{4\pi} \left(F^{\mu\nu}F_{\mu\nu}g^{\alpha\beta}+d\, F^{\alpha\gamma}F_{\gamma}^{ \beta}\right) \Bigg\rangle^{\mu_3\mu_4} \, .
\end{align}
In the final step we recognised the photon field strength matrix element. 
The $\overline{\mathrm{MS}}$ prescription consists in removing the $\frac{1}{\hat{\epsilon}}$ pole from the matrix element and incorporating a scale $\mu$ to make the implicit logarithm in $Q_{3}^{-2\epsilon}$ dimensionless, i.e.
\begin{align}\nonumber
&\lim_{q_4 \rightarrow 0}\partial^{\nu_3}_{q_3}\, \partial^{\nu_4}_{q_4}\, \Bigg\langle \bar{q}(0)\left[ \overrightarrow{D}^{\alpha}-\overleftarrow{D}^{\alpha}\right] \gamma^{\beta}  q(0) \Bigg\rangle ^{\mu_3,\, \mu_4}_{\overline{\mathrm{MS}}(\mu)}
\\
&
=
{\textrm{finite }}+i\, \frac{2}{3} \, N_c\, e_{q,j}^2\, \ln\frac{Q_3^2}{\mu^2}\lim_{q_4 \rightarrow 0}\partial^{\nu_3}_{q_3}\partial^{\nu_4}_{q_4}\Bigg\langle  \frac{\alpha}{4\pi} \left( F^{\mu\nu}F_{\mu\nu}g^{\alpha\beta}+d\, F^{\alpha\gamma}F_{\gamma}^{ \beta}\right) \Bigg\rangle^{\mu_3\mu_4} 
\, .
\end{align}
Writing $\mathcal{O}^{\alpha\beta}_j=\bar{q}_j(0)\left[ \overrightarrow{D}^{\alpha}-\overleftarrow{D}^{\alpha}\right] \gamma^{\beta}  q_j(0)$ and denoting the bare operator by $\mathcal{O}^{\alpha\beta}_{(0),\, j}$, the mixing is of the form
\begin{equation}
\mathcal{O}^{\alpha\beta}_{(0),\, j}=\mathcal{O}^{\alpha\beta}_j(\mu)+Z^j_{DF}(\mu)\, \frac{\alpha}{4\pi}\, \left(F^{\mu\nu}F_{\mu\nu}g^{\alpha\beta}+d\, F^{\alpha\gamma}F_{\gamma}^{\beta}\right) +\ldots
\, ,
\end{equation}
where the renormalization factor $Z^j_{DF}(\mu)$ is
\begin{equation}
Z^j_{DF}(\mu)=-i\, \frac{2}{3}\,  N_c \, e_{q,j}^2\, \frac{\mu^{-2\epsilon}}{\hat{\epsilon}} \, .
\end{equation}
Repeating exactly the same calculation but with two initial-state gluons we find
\begin{equation}
\mathcal{O}^{\alpha\beta}_{(0),\, j}=\mathcal{O}^{\alpha\beta}_j(\mu)+Z^j_{DG}(\mu)\, \frac{\alpha_s}{4\pi}\, \left(
G^{a,\mu\nu}G^{a}_{\mu\nu}\, g^{\alpha\beta}+d\, G^{\alpha\gamma}_a G_{\gamma }^{a,\, \beta}
\right) 
+\ldots
\, , 
\end{equation}
with
\begin{equation}
Z^j_{DG}(\mu)=-i\, \frac{1}{3}\,  \frac{\mu^{-2\epsilon}}{\hat{\epsilon}} \, .
\end{equation}
These are needed to have finite Wilson coefficients associated to the field strength contributions and quark-current matrix elements at the associated leading order contributions in Eq.~(\ref{eq:masterope}).

The renormalized version of Eq.~\eqref{eq:masterope} is then
\begin{small}
\begin{align}\nonumber
&\lim_{q_4\rightarrow 0} \frac{\partial \Pi^{\mu_1\mu_2\mu_3\nu_4}}{\partial q_{4}^{\mu_4}}
=
\sum_j\frac{ e_{q,j}^2}{e^2}\,   \lim_{q_4 \rightarrow 0}\partial ^{\nu_4}_{q_4}\, \Bigg\langle\bar{q}(0)\left[
\Gamma^{\mu_1\mu_2}(-\hat{q})-\Gamma^{\mu_2\mu_1}(-\hat{q})
\right]
q(0) \Bigg\rangle^{j, \mu_3,\mu_4} 
\\
&
+\sum_j\frac{ie_{q_j}^2}{e^2\hat{q}^2}\, 
\left(g^{\mu_1\delta}g^{\mu_2}_{\beta}+g^{\mu_2\delta}g^{\mu_1}_{\beta}-g^{\mu_1\mu_2}g^{\delta}_{\beta}\right)
\left(g_{\alpha}^{\delta}-2\, \frac{\hat{q}^\delta \hat{q}_\alpha}{\hat{q}^2}\right)
\nonumber \\
& 
\qquad 
\times 
\lim_{q_4 \rightarrow 0}\partial_{\nu_4}^{q_4}\, \Bigg\langle \bar{q}(0)(\overrightarrow{D}^{\alpha}-\overleftarrow{D}^{\alpha})\gamma^{\beta}  q(0) \Bigg\rangle ^{j, \mu_3,\, \mu_4}_{\overline{\mathrm{MS}}(\mu)} \nonumber \\
\nonumber \\
&
+\sum_j\frac{ie_{q_j}^2}{e^2\hat{q}^2}\, 
\left(g^{\mu_1\delta}g^{\mu_2}_{\beta}+g^{\mu_2\delta}g^{\mu_1}_{\beta}-g^{\mu_1\mu_2}g^{\delta}_{\beta}\right)
\left(g_{\alpha}^{\delta}-2\, \frac{\hat{q}^\delta \hat{q}_\alpha}{\hat{q}^2}\right) 
\nonumber \\
&
\qquad
\times 
\lim_{q_4 \rightarrow 0}\partial^{\nu_4}_{q_4} \, 
\Bigg\langle
Z^j_{DF}(\mu)\, \frac{\alpha}{4\pi}\, \left(F^{\mu\nu}F_{\mu\nu}g^{\alpha\beta}+d\, F^{\alpha\gamma}F_{\gamma}^{ \beta}\right)
\nonumber \\
& \qquad \qquad \qquad
+ Z^j_{DG}(\mu)\, \frac{\alpha_s}{4\pi}\, \left( G^{\mu\nu}_a G_{\mu\nu}^a \,  g^{\alpha\beta}+d\, G^{\alpha\gamma}_a G_{\gamma}^{a , \, \beta}\right) 
\Bigg\rangle^{j,\mu_3\mu_4}
\nonumber \\
&
+ \sum_j   \frac{ e_{q_j}^{2}}{8 e^2} \lim_{q_4 \rightarrow 0} \left[ \partial^{\nu_4}_{q_4}\, \Bigg\langle e^2 e_{q_j}^2  F_{\nu_3'\mu_3'}F_{\nu_4'\mu_4'} +\frac{1}{2N_{c}}\,  g_s^2 \,  G^{a}_{\nu_3'\mu_3'}G^{a}_ {\nu_4'\mu_4'} \Bigg\rangle^{j,\mu_3\mu_4}\right] \, 
\lim_{q_3,q_4 \rightarrow 0}\partial_{q_3}^{\nu_3'}\, \partial_{q_4}^{\nu_4'}\, \Pi_{\mathrm{ql},j}^{\mu_1\mu_2\mu_3 '\mu_4 '}
\, , \label{eq:masteroperen}
\end{align}
\end{small}
The last four lines together yield a renormalized Wilson coefficient. Within this renormalization prescription the field strength contributions
to the $\hat{\Pi}_i $ are found to be zero, except for
\begin{align}\label{eq:pihatff}
\underline{q_{3}\textrm{ small:}} &
\nonumber \\
\hat{\Pi}_4 & =-\frac{16}{3\pi^2 \overline{Q}_3^4} \, ,   
\nonumber \\
\underline{q_{1}\textrm{ small:}} &
\nonumber \\
 \hat{\Pi}_1& =\frac{8}{3\pi^2 \overline{Q}_1^4}\left(1-4\log\frac{\overline{Q}_1}{2\mu}\right) \, ,
\nonumber \\
\underline{q_{2}\textrm{ small:}} &
\nonumber \\
 \hat{\Pi}_1 & =\frac{8}{3\pi^2 \overline{Q}_2^4}\left(1-4\log\frac{\overline{Q}_2}{2\mu}\right) \, .   
\end{align}
This can be compared to the non-renormalized case in Eqs.~(\ref{eq:d4photonq3}--\ref{eq:d4photonq2}). The remaining $\hat{\Pi}_i$ contributions are straightforward to obtain by simply subtracting this contribution and the ones coming from $D=3$ to the total $\hat{\Pi}_i$ in Eqs.~(\ref{eq:d4pihatfinalq3}--\ref{eq:d4pihatfinalq2}). 
As a final remark, to the order expanded the OPE in Eq.~(\ref{eq:masteroperen}) satisfies gauge invariance in both $q_1$ and $q_2$ as well.

\section{Nonperturbative regime}
\label{sec:nonperturbative}

In this section we extend the results of Sec.~\ref{sec:alphaSLO} beyond the $Q_{3}^2\gg \Lambda_{\mathrm{QCD}}^2 $ regime. We start by reproducing the $D=3$ limit of Ref.~\cite{Melnikov:2003xd}. For the power-suppressed $D=4$ extension we directly work in the chiral limit, where we conveniently separate the tensor of Eq.~(\ref{eq:tensorbracket}) in two contributions, a flavor singlet $\Pi^{\mu_1\mu_2\mu_3\mu_4}_{(1)}$ and a flavor octet $\Pi^{\mu_1\mu_2\mu_3\mu_4}_{(8)}$, according to
\begin{align}\nonumber
\Pi^{\mu_1\mu_2\mu_3\mu_4}
&
=\sum_j \frac{i e_{q_j}^{2}}{e^2} \Bigg\langle e^{-i(q_1 x_1 +q_2 x_2)}\, \bar{q}(x_1)\gamma^{\mu_1}q(x_1) \;\bar{q}(x_2)\gamma^{\mu_2}q(x_2) \Bigg\rangle ^{j,\mu_3,\, \mu_4}_{q_{4},\, x _1,\, x_2} 
\, \\ \nonumber
&= \frac{i \sum_j (e_{q_j}^{2}-\sum_{k} \frac{e_{q_k}^2}{3})}{e^2}\, 
\Bigg\langle e^{-i(q_1 x_1 +q_2 x_2)}\, \bar{q}(x_1)\gamma^{\mu_1}q(x_1) \;\bar{q}(x_2)\gamma^{\mu_2}q(x_2) \Bigg\rangle ^{j,\mu_3,\, \mu_4}_{q_{4},\, x _1,\, x_2} 
\\
&+\frac{i \sum_{j}\sum_{k} \frac{e_{q_k}^2}{3}}{e^2}\Bigg\langle e^{-i(q_1 x_1 +q_2 x_2)}\, 
\bar{q}(x_1)\gamma^{\mu_1}q(x_1) \;\bar{q}(x_2)\gamma^{\mu_2}q(x_2) \Bigg\rangle ^{j,\mu_3,\, \mu_4}_{q_{4},\, x _1,\, x_2}\nonumber \\
&\equiv \Pi^{\mu_1\mu_2\mu_3\mu_4}_{(8)}+\Pi^{\mu_1\mu_2\mu_3\mu_4}_{(1)} \, .
\end{align}
Obviously, we will have a much better predictive power for the octet case because all gluon operator contributions (independent of $j$) cancel, but the form of the obtained OPE is also valid for the singlet. 

All $\hat{\Pi}_i$ and form factors are in this section given in units of $N_{c}\sum_j e_{q_j}^4$.

\subsection{Dimension $D=3$}

Let us start by rewriting the dimension $D=3$ part of Eq.~(\ref{eq:masterope}) as
\begin{align}\label{eq:dim3gen}
\lim_{q_4 \rightarrow 0}\frac{\partial\Pi^{\mu_1\mu_2\mu_3 \nu_4}}{\partial q_{4,\, \mu_4}}
&
=
-\sum_{j}\frac{e_{q,j}^2}{e^2}\, \frac{\hat{q}^{\alpha}}{\hat{q}^2}
\, 
g^{\alpha_1\mu_{1}}g^{\alpha_2\mu_2}
\nonumber \\
& 
\times \lim_{q_4 \rightarrow 0}\partial_{q_4}^{\nu_4}\,  \langle  0|\bar{q}_{j} (\gamma_{\alpha_1}\gamma_{\alpha} \gamma_{\alpha_2}-\gamma_{\alpha_2}\gamma_{\alpha} \gamma_{\alpha_1}) q_{j}|\gamma^{\mu_3}(q_3)\gamma^{\mu_4}(q_4)\rangle \, .
\end{align}
Using the relation\footnote{We work with the convention $\epsilon^{0123}=1$.}
\begin{equation}
\gamma_{\alpha_1}\gamma_{\alpha}\gamma_{\alpha_2}-\gamma_{\alpha_2}\gamma_{\alpha}\gamma_{\alpha_1}=-2i\, \epsilon_{\delta\alpha_1\alpha\alpha_2}\gamma^\delta \gamma_5 \, ,
\end{equation}
one finds
\begin{align}
\lim_{q_4 \rightarrow 0}\frac{\partial\Pi^{\mu_1\mu_2\mu_3 \nu_4}}{\partial q_{4,\, \mu_4}}
&=2\sum_{j}\frac{e_{q,j}^2}{e^2}\,\epsilon^{\mu_1\mu_2\alpha\delta}\, \frac{\hat{q}^{\alpha}}{\hat{q}^2}\lim_{q_4 \rightarrow 0}i\, \partial_{q_4}^{\nu_4}\, \langle  0|\bar{q}_{j} \gamma^{\delta}\gamma_5 q_{j}|\gamma^{\mu_3}(q_3)\gamma^{\mu_4}(q_4)\rangle .
\end{align}
In the nonperturbative regime without an inherent ordering on $Q_{3}$ and $\Lambda _{\textrm{QCD}}$, we must express the matrix element in a form factor decomposition. The most general decomposition into Lorentz structures and associated form factors is given by
\begin{align}\nonumber
&\sum_{j}\frac{e_{q,j}^2}{e^2}\lim_{q_4 \rightarrow 0}\, i\, \partial_{q_4}^{\nu_4}\langle  0|\bar{q}_{j} \gamma^{\delta}\gamma_5 q_{j}|\gamma^{\mu_3}(q_3)\gamma^{\mu_4}(q_4)\rangle
\\
&=
\frac{1}{4\pi^2}\, q_{3}^2\, \left(\epsilon_{\mu_3\mu_4\nu_4\delta}\, \omega_{T}(q_3^{2})
-\frac{1}{q_3^2}\, \epsilon_{q_3\mu_4\nu_4\delta}\, q_{3\mu_3}\, \omega_{T}(q_3^{2})
+\frac{1}{q_3^2}\, \epsilon_{\mu_3 \mu_4 \nu_4 q_3}\, q_{3\delta}\, \left[\omega_{L}(q_3^2)-\omega_{T}(q_3^2)\right] \right) \, .
\end{align}
Here we introduced the compact notation $\epsilon_{\cdots q \cdots}\equiv \epsilon_{\cdots \mu \cdots}q^{\mu}$, as well as the longitudinal and transversal form factors $\omega _{L,T}(q_3^2)$ following the notation in Ref.~\cite{Melnikov:2003xd}. This yields the HLbL derivative tensor as
\begin{align}
\lim_{q_4 \rightarrow 0} & \frac{\partial\Pi^{\mu_1\mu_2\mu_3  \nu_4}}{\partial q_{4,\, \mu_4}}
=\frac{1}{2\pi^2}\, \frac{q_{3}^2}{\hat{q}^2}
\, \epsilon^{\mu_1\mu_2\hat{q}\delta}
\Big(\epsilon_{\mu_3\mu_4\nu_4\delta}\, \omega_{T}(q_3^{2})-\frac{1}{q_3^2}\, \epsilon_{q_3\mu_4\nu_4\delta}\, q_{3\mu_3}\, \omega_{T}(q_3^{2})
\nonumber \\
&
\hspace{27.5ex}
+\frac{1}{q_3^2}\, \epsilon_{\mu_3 \mu_4 \nu_4 q_3}\, q_{3\delta}\, \left[ \omega_{L}(q_3^2)-\omega_{T}(q_3^2)\right]\Big) \, .
\end{align}
When $Q_3^2$ is large, i.e.~in the perturbative regime, the partonic description holds and
\begin{equation}\label{eq:partD3}
\omega_{L}=2\, \omega_{T}=\frac{2}{-Q_{3}^2} \, ,
\end{equation}
which agrees with the $D=3$ result obtained in Sec.~\ref{sec:alphaSLO} and with Ref.~\cite{Melnikov:2003xd}. Next, we can match to find the corresponding generalized $\hat{\Pi}_i$. In the $q_{3}$ small region one finds at LO in the $\frac{1}{\overline{Q}_3}$ counting,
\begin{align}
\underline{q_{3}\textrm{ small:}} &
\nonumber \\
\hat{\Pi}_{1}& =\frac{2}{\pi^2\overline{Q}_{3}^2}\, \omega_{L}(q_3^2) \, .
\end{align}
This can be compared to Eq.~(\ref{eq:d3quarkq3}). 
The partial contribution coming from $D=3$ at NLO in the $\frac{1}{\overline{Q}_3}$ counting, i.e.~extending Eq.~(\ref{eq:d3expquarkq3}), is given by 
\begin{align}
\underline{q_{3}\textrm{ small:}} &
\nonumber \\
\hat{\Pi}_{17} & =-\frac{4 Q_3^2}{\pi^2(Q_3^2-\delta_{12}^2)\overline{Q}_3^4}\, \omega_T(q_3^2)\, ,
\nonumber \\
\hat{\Pi}_{39}& =-\frac{4 Q_3^2}{\pi^2(Q_3^2-\delta_{12}^2)\overline{Q}_3^4}\, \omega_T(q_3^2)\, .
\end{align}
Through NLO in the $q_{1}$ small region we find
\begin{align}
\underline{q_{1}\textrm{ small:}} &
\nonumber \\
\hat{\Pi}_{4}&=\frac{ Q_1^2}{\pi^2(Q_1^2-\delta_{23}^2)\overline{Q}_1^2}\, \omega_T(q_1^2)\, ,\\
\hat{\Pi}_{7}&=-\frac{4 Q_1^2 \delta_{23}}{\pi^2(Q_1^2-\delta_{23}^2)\overline{Q}_1^3}\, \omega_T(q_1^2)\, ,\\
\hat{\Pi}_{39}&=-\frac{4 Q_1^2 }{\pi^2(Q_1^2-\delta_{23}^2)\overline{Q}_1^4}\, \omega_T(q_1^2)\, ,\\
\hat{\Pi}_{54}&=-\frac{4 Q_1^2 }{\pi^2(Q_1^2-\delta_{23}^2)\overline{Q}_1^4}\, \omega_T(q_1^2)\, .
\end{align}
Finally in the $q_2$ small region the result is
\begin{align}
\underline{q_{2}\textrm{ small:}} &
\nonumber \\
\hat{\Pi}_{4}&=\frac{ Q_2^2}{\pi^2(Q_2^2-\delta_{31}^2)\overline{Q}_2^2}\, \omega_T(q_2^2)\, ,\\
\hat{\Pi}_{7}&=\frac{4 Q_2^2}{\pi^2(Q_2^2-\delta_{31}^2)\overline{Q}_2^4}\, \omega_T(q_2^2)\, ,\\
\hat{\Pi}_{39}&=-\frac{4 Q_2^2}{\pi^2(Q_2^2-\delta_{31}^2)\overline{Q}_2^4}\, \omega_T(q_2^2)\, ,\\
\hat{\Pi}_{54}&=\frac{4 Q_2^2}{\pi^2(Q_2^2-\delta_{31}^2)\overline{Q}_2^4}\, \omega_T(q_2^2)\, .
\end{align}

From the collinear divergences $\delta _{ij}\rightarrow Q_k$ in these expressions it is once again transparent how keeping only the $D=3$ at NLO in $\frac{1}{\overline{Q}_k}$ is not consistent: the $\hat{\Pi}_i$ are by construction free from kinematic singularities while these partial results are not. It can easily be checked that by saturating the form factors to their short-distance limits of Eq.~(\ref{eq:partD3}), one recovers the corresponding $\hat{\Pi}_i$ in Eqs.~(\ref{eq:d3expquarkq3}--\ref{eq:d3expquarkq2}).

Even if it is not the focus of this work and there are many of them about the topic ~\cite{Knecht:2002hr,Vainshtein:2002nv,Czarnecki:2002nt,Knecht:2003xy,Knecht:2020xyr}, let us note how in the chiral limit further simplifications can be made. First, the third and fourth photon must emerge from the same quark line (and thus from the same flavor), otherwise one would have a vanishing flavor trace, $\mathrm{Tr}(e_{q})=0$. We then have a disconnected and a connected topology. For the same reason, the octet piece $\Pi_{(8)}^{\mu_1 \mu_2 \mu_3 \mu_4}$, does not receive any contribution from disconnected topologies.

The octet form factors, $\omega_{L,(8)}$ and $\omega_{T,(8)}$, are the ones satisfying nonrenormalization theorems \cite{Vainshtein:2002nv,Melnikov:2003xd,Knecht:2001qf}. Concretely, the longitudinal octet value given by 
\begin{align}
\omega_{L,(8)}=\frac{\sum_{j}e_{q,j}^4-\frac{1}{3}\sum_{j,k}e_{q,j}^2e_{q,k}^2}{\sum_j e_{q,j}^4}\omega_L=\frac{1}{3}\omega_{L} \, ,
\end{align}
from Eq.~(\ref{eq:partD3})  is exact beyond perturbation theory, i.e.~for all $Q_3^2$ values, and $\omega_{T,(8)}$ does not receive any perturbative correction. However, the validity of Eq.~(\ref{eq:partD3}) cannot be extended beyond perturbation theory, i.e. as $Q_{3}^2\rightarrow \Lambda_{\mathrm{QCD}}^2$. While much less can be said about the singlet, one naively expects its contribution to be suppressed with respect to the octet because there are no zero mass modes in the singlet axial channel.

\subsection{Dimension $D=4$}

As explained above, for the suppressed $D=4$ contribution we directly work in the chiral limit. In Table~\ref{tab:D4cont} we summarise which contributions appear at dimension $D=4$. This follows since the disconnected topology vanishes for $\Pi_{(8)}^{\mu_1 \mu_2 \mu_3 \mu _4}$ and that the gluon operator and the photon one (at this order in $\alpha_s(\overline{Q}_3)$) respectively are disconnected and connected.
\begin{table}[tb]\centering
\begin{tabular}{|c|cc|cc|}
\hline
\multirow{2}{*}{Operator} & \multicolumn{2}{c|}{$\Pi_{(1)}$}                       & \multicolumn{2}{c|}{$\Pi_{(8)}$}                       \\ \cline{2-5} 
                          & \multicolumn{1}{c|}{Connected}       & Disconnected    & \multicolumn{1}{c|}{Connected}       & Disconnected    \\ \hline
Photon                     & \multicolumn{1}{c|}{$\newcheckmark$} & $\newcrossmark$ & \multicolumn{1}{c|}{$\newcheckmark$} & $\newcrossmark$ \\ \hline
Gluon                    & \multicolumn{1}{c|}{$\newcrossmark$} & $\newcheckmark$ & \multicolumn{1}{c|}{$\newcrossmark$} & $\newcrossmark$ \\ \hline
Quark                     & \multicolumn{1}{c|}{$\newcheckmark$} & $\newcheckmark$ & \multicolumn{1}{c|}{$\newcheckmark$} & $\newcrossmark$ \\ \hline
\end{tabular}
\caption{\label{tab:D4cont}Summary of studied contributions entering at dimension $D=4$.}
\end{table}

\subsubsection{Octet case}
For the octet case the gluon operator contribution trivially drops due to its flavorless nature. The photon matrix element does not receive nonperturbative correction at the studied order in $\alpha$ and it is therefore identical to the contribution studied in Sec.~\ref{sec:alphaSLO}, except for a trivial charge factor. The associated NLO contribution to the OPE then becomes
\begin{align}\nonumber
&\lim_{q_4\rightarrow 0} \frac{\partial \Pi^{\mu_1\mu_2\mu_3\nu_4}_{(8)}}{\partial q_{4,\, \mu_4}}
= 
\frac{1}{3}\frac{\partial \Pi^{\mu_1\mu_2\mu_3\nu_4}_{FF,\mu}}{\partial q_{4,\, \mu_4}}
\\
&
+\frac{i\sum_j \left(e_{q_j}^2 \nonumber -\sum_k\frac{e_{q,k}^2}{3}\right)}{e^2\hat{q}^2}\, 
\left( g^{\mu_1}_{\delta} g^{\mu_2}_{\beta}+g^{\mu_2}_{\delta}g^{\mu_1}_{\beta}-g^{\mu_1\mu_2}g_{\delta\beta}\right)
\left(g_{\alpha}^{\delta}-2\, \frac{\hat{q}^\delta \hat{q}_\alpha}{\hat{q}^2}\right) 
\nonumber \\
&
\times 
\lim_{q_4 \rightarrow 0}\partial^{\nu_4}_{q_4}\, \Bigg\langle \bar{q}(0)\left[ \overrightarrow{D}^{\alpha}-\overleftarrow{D}^{\alpha}\right] \gamma^{\beta}  q(0) \Bigg\rangle ^{j, \mu_3,\, \mu_4}_{\overline{\mathrm{MS}}(\mu)} \nonumber  \, .
\end{align}
where $\Pi^{\mu_1\mu_2\mu_3\nu_4}_{FF,\mu}$ simply refers to the Wilson-renormalized photon operator contribution in Eq.~(\ref{eq:masteroperen}), whose contribution to the $\hat{\Pi}_i$ is still given by Eq.~(\ref{eq:pihatff}). The remaining matrix element with 5 Lorentz indices $\alpha \beta \mu_3 \mu_4 \nu_4$ can be decomposed into form factors. There are in total 26 possible Lorentz structures from combinations of $q_3^\mu$ and $g^{\nu\rho}$ (the Levi-Civita tensor does not appear because of parity). However, we know that the final matrix element has to be anti-symmetric in $\mu_4 \leftrightarrow \nu _4$ and be gauge invariant with respect to $q_{3}^{\mu_3}$. Applying these conditions yields a set of 6 independent form factors $\omega_{(8)}^{D,i}(q_3^2)$ so that the matrix element can be written
\begin{equation}
\frac{i\sum_j \left(e_{q_j}^2  -\sum_k\frac{e_{q,k}^2}{3}\right)}{e^2\hat{q}^2}  
\lim_{q_4 \rightarrow 0}\, \partial ^{\nu_4}_{q_4} \, \Bigg\langle \bar{q}(0)\left[ \overrightarrow{D}^{\alpha}-\overleftarrow{D}^{\alpha}\right]\gamma^{\beta}  q(0) \Bigg\rangle ^{j, \mu_3,\, \mu_4}_{\overline{\textrm{MS}}(\mu)}=\sum_{i=1}^6 \omega_{(8)}^{D,i}\, L_i^{\alpha\beta\mu_3\mu_4\nu_4} \, . \label{eq:octetME}
\end{equation}
The Lorentz structures are
\begin{align}\nonumber
L^{\alpha\beta\mu_3\mu_4\nu_4}_1&=g^{\mu_3\mu_4} q_3^{\nu_4} g^{\alpha\beta}-(\mu_4 \leftrightarrow \nu_4)
\, ,          \\ \nonumber
L^{\alpha\beta\mu_3\mu_4\nu_4}_2&=g^{\beta \mu_4}q_{3}^{\nu_4}\left(g^{\alpha\mu_3} -\frac{q_3^{\alpha} q_3^{\mu_3}}{q_3^2}\right)- (\mu_4 \leftrightarrow \nu_4) \, ,\\ \nonumber
L^{\alpha\beta\mu_3\mu_4\nu_4}_3&=L^{\beta\alpha\mu_3\mu_4\nu_4}_2 \, ,\\ \nonumber
L^{\alpha\beta\mu_3\mu_4\nu_4}_4&=g^{\alpha\mu_4}q_3^{\beta}\left(g^{\mu_3\nu_4} -\frac{q_3^{\mu_3} q_3^{\nu_4}}{q_3^2}\right) - (\mu_4 \leftrightarrow \nu_4) \, ,\\ \nonumber
L^{\alpha\beta\mu_3\mu_4\nu_4}_5&=L^{\beta\alpha\mu_3\mu_4\nu_4}_4 \, ,\\
L^{\alpha\beta\mu_3\mu_4\nu_4}_6&=g^{\mu_3\mu_4} q_3^{\nu_4} q^{\alpha}_3 q^{\beta}_3-(\mu_4 \leftrightarrow \nu_4)  \, .
\end{align}

Projecting onto the $\hat{\Pi}_i$ and requiring that they are free of kinematic singularities, we obtain nontrivial relations among nonperturbative form factors.
The conditions are\footnote{Let us note how the $\omega_{(8)}^{D,4}=\omega_{(8)}^{D,5}$ and the $\omega_{(8)}^{D,3}-\omega_{(8)}^{D,2}=\frac{Q_i^2}{4\pi^2}\omega_{T,(8)}$ relations are constraints on the antisymmetric part of the quark operator, which at the studied order does not mix with the photon one.}
\begin{align}
    \omega _{(8)}^{D,2} & = -2\, \omega _{(8)}^{D,1}+ \omega _{(8)}^{D,5} -\frac{\omega _{(8)}^{D,6}}{2}-\frac{\omega _{T,(8)}Q_i^2}{8\pi^2 } \, ,
    \nonumber \\
    \omega _{(8)}^{D,3} & = -2\, \omega _{(8)}^{D,1}+ \omega _{(8)}^{D,5} -\frac{\omega _{(8)}^{D,6}}{2}+\frac{\omega _{T,(8)}Q_i^2}{8\pi^2 }
    \, ,
    \nonumber \\
    \omega _{(8)}^{D,4} & =\omega _{(8)}^{D,5} \, . 
\end{align}
Using these relations, we find the $\hat{\Pi}_{i}$ to be
\begin{align}
\underline{q_{3}\textrm{ small:}} &
\nonumber \\
\hat{\Pi}_1&=\mathcal{O}\left( \frac{1}{\overline{Q}_{3}^4}\right) \, ,\\
\hat{\Pi}_4&=-\frac{64\left( \omega_{(8)}^{D,1}-\omega_{(8)}^{D,5}\right)  }{\overline{Q}_{3}^4} 
\, ,\\
\hat{\Pi}_7
&
=
\mathcal{O}\left( \frac{1}{\overline{Q}_{3}^6}\right)
\, ,\\
\hat{\Pi}_{17}
&=\frac{32 \left(-2 \omega_{(8)}^{D,1}+2\omega_{(8)}^{D,5}+\frac{\omega_{(8)}^{D,6}}{2}-\frac{\omega_{T}Q_3^2}{8\pi^2}\right) }{\overline{Q}_{3}^4 Q_{3}^2} \, ,
\, ,\\
\hat{\Pi}_{39}&=-\frac{32 \left(-2 \omega_{(8)}^{D,1}+2\omega_{(8)}^{D,5}-\frac{\omega_{(8)}^{D,6}}{2}+\frac{\omega_{T}Q_3^2}{8\pi^2}\right) }{\overline{Q}_{3}^4 Q_{3}^2} \, ,\\
\hat{\Pi}_{54}&=\mathcal{O}\left( \frac{1}{\overline{Q}_{3}^5}\right) \, .
\end{align}
\begin{align}
\underline{q_{1}\textrm{ small:}} &
\nonumber \\
\hat{\Pi}_1&=\frac{32\, \omega_{(8)}^{D,5} }{\overline{Q}_{1}^4} \, ,\\
\hat{\Pi}_4&=-\frac{ 
16\,   \omega_{(8)}^{D,1}
-16\,  \omega_{(8)}^{D,5}
+4 \, \omega _{(8)}^{D,6} - \frac{\omega _{T,(8)}Q_1^2}{\pi^2}
}{Q_{1}^2 \overline{Q}_{1}^2}
\, ,\\
\hat{\Pi}_7&=\mathcal{O}\left(\frac{1}{\overline{Q}_{1}^4} \right) \, ,\\
\hat{\Pi}_{17}&=\mathcal{O}\left(\frac{1}{\overline{Q}_{1}^5} \right) \, ,\\
\hat{\Pi}_{39}&=\frac{4 \left(
16\,   \omega_{(8)}^{D,1}
-16\,  \omega_{(8)}^{D,5}
+4 \omega _{(8)}^{D,6} - \frac{\omega _{T,(8)}Q_1^2}{\pi^2}
\right)
}{Q_{1}^2 \overline{Q}_{1}^4}  \, ,\\
\hat{\Pi}_{54}&=-\frac{4 \left(
16\,   \omega_{(8)}^{D,1}
-16\,  \omega_{(8)}^{D,5}
+4 \, \omega _{(8)}^{D,6} 
+ \frac{\omega _{T,(8)}Q_1^2}{\pi^2}
\right)
}{Q_{1}^2 \overline{Q}_{1}^4} \, .
\end{align}
\begin{align}
\underline{q_{2}\textrm{ small:}} &
\nonumber \\
\hat{\Pi}_1&=\frac{32}{\overline{Q}_{2}^4}\, \omega_{(8)}^{D,5} \, ,\\
\hat{\Pi}_4&=-\frac{ 
16\,   \omega_{(8)}^{D,1}
-16\,  \omega_{(8)}^{D,5}
+4 \, \omega _{(8)}^{D,6} 
- \frac{\omega _{T,(8)}Q_2^2}{\pi^2}
}{Q_{2}^2 \overline{Q}_{2}^2}
\, ,\\
\hat{\Pi}_7&=-4\frac{ 
16\,   \omega_{(8)}^{D,1}
-16\,  \omega_{(8)}^{D,5}
+4 \, \omega _{(8)}^{D,6} 
- \frac{\omega _{T,(8)}Q_2^2}{\pi^2}
}{Q_{2}^2 \overline{Q}_{2}^4}
\, ,\\
\hat{\Pi}_{17}&=\mathcal{O}\left(\frac{1}{\overline{Q}_{2}^5} \right) \, ,\\
\hat{\Pi}_{39}&=4\frac{ 
16\,   \omega_{(8)}^{D,1}
-16\,  \omega_{(8)}^{D,5}
+4 \, \omega _{(8)}^{D,6} 
- \frac{\omega _{T,(8)}Q_2^2}{\pi^2}
}{Q_{2}^2 \overline{Q}_{2}^4} \, ,\\
\hat{\Pi}_{54}&=4\frac{ 
16\,   \omega_{(8)}^{D,1}
-16\,  \omega_{(8)}^{D,5}
+4 \, \omega _{(8)}^{D,6} 
+ \frac{\omega _{T,(8)}Q_2^2}{\pi^2}
}{Q_{2}^2 \overline{Q}_{2}^4} \, .
\end{align}
In fact, me may here take the perturbative limits $Q_i^2\gg \Lambda _{\textrm{QCD}}^2$ and match the three appearing form factors $\omega _{(8)}^{D,1}$, $\omega _{(8)}^{D,5}$ and $\omega _{(8)}^{D,6}$. Using Eq.~(\ref{eq:partD3}), the result is
\begin{align}\label{eq:octetpertlimit}
    \underline{q_{i}\textrm{ small:}} &
\nonumber \\
\omega _{(8)}^{D,1} 
&
= \frac{-13+12 \log \frac{Q_i}{\mu}}{108\, \pi^2}
\, ,
\nonumber \\
\omega _{(8)}^{D,5} 
&
= \frac{-13+12 \log \frac{Q_i}{\mu}}{108\, \pi^2}
\, ,
\nonumber \\
\omega _{(8)}^{D,6} 
&
= \frac{1}{36\, \pi^2}
\, .
\end{align}

The full NLO octet result at the nonperturbative level becomes then parameterized in terms of $3$ form factors depending on a single energy scale, whose short-distance dependence is known and with a clear definition, given by Eq.~(\ref{eq:octetME}). Complementing this model-independent result with nonperturbative methods to extrapolate them to lower energies, such as resonance models, should allow for a more reliable estimate of the subleading contributions to $g-2$ coming from this particularly challenging domain. 

\subsubsection{Singlet case}
More complicated is the singlet case, since the gluon operator does not drop out. In this case the NLO piece takes the form

\begin{align}\nonumber
&\lim_{q_4\rightarrow 0}  \frac{\partial \Pi^{\mu_1\mu_2\mu_3\nu_4}_{(1)}}{\partial q_{4,\, \mu_4}}
= \frac{2}{3}\frac{\partial \Pi^{\mu_1\mu_2\mu_3\nu_4}_{FF,\mu}}{\partial q_{4}^{\mu_4}}
\\
&
+i\, \frac{ \sum_j\sum_k \frac{e_k^2}{3}}{e^2\hat{q}^2} \, 
\left( g^{\mu_1}_{\delta}g^{\mu_2}_{\beta}+g^{\mu_2}_{\delta}g^{\mu_1}_{\beta}-g^{\mu_1\mu_2}g_{\delta\beta}\right)
\left(g_{\alpha}^{\delta}-2\, \frac{\hat{q}^\delta \hat{q}_\alpha}{\hat{q}^2}\right) 
\nonumber \\
&
\qquad
\times
\lim_{q_4 \rightarrow 0}\partial ^{\nu_4}_{q_4}\, \Bigg\langle \bar{q}(0)\left[ \overrightarrow{D}^{\alpha}-\overleftarrow{D}^{\alpha}\right]\gamma^{\beta}  q(0) \Bigg\rangle ^{j, \mu_3,\, \mu_4}_{ \overline{\textrm{MS}}(\mu)} \nonumber \\
&
+\frac{i\, \sum _j \sum_k \frac{e_{q_k}^2}{3}}{e^2\hat{q}^2}\left( g^{\mu_1}_{\delta}g^{\mu_2}_{\beta}+g^{\mu_2}_{\delta}g^{\mu_1}_{\beta}-g^{\mu_1\mu_2}g_{\delta\beta}\right)
\left(g_{\alpha}^{\delta}-2\, \frac{\hat{q}^\delta \hat{q}_\alpha}{\hat{q}^2}\right) 
\nonumber \\
&
\qquad
\times \lim_{q_4 \rightarrow 0}\partial^{\nu_4}_{q_4}\,  \left[ Z^j_{DG}(\mu)\, \frac{\alpha_s}{4\pi}\, \left(G^{\mu\nu}_a G_{\mu\nu}^a \, g^{\alpha\beta}+d\, G^{\alpha\gamma}_a G^{a,\, \beta}_{\gamma}\right)\right]  \nonumber\\
&+
\frac{ \sum_j \sum_k \frac{e_{q_k}^{2}}{3}}{8 e^2}\,  \lim_{q_4 \rightarrow 0}\partial^{\nu_4}_{q_4}\, \left[
\Bigg\langle \frac{1}{2N_{c}} \, g_s^2 \, G^{a}_{\nu_3'\mu_3'}G^{a}_{\nu_4'\mu_4'} \Bigg\rangle^{j,\mu_3\mu_4}\right] \,
\times \lim_{q_3,q_4 \rightarrow 0}\partial_{q_3}^{\nu_3'}\partial_{q_4}^{\nu_4'}\Pi_{\mathrm{ql,j}}^{\mu_1\mu_2\mu_3 ' \mu_4 '}
\, .
\end{align}

We may decompose the renormalized quark-current matrix element in precisely the same way, only now with 6 singlet form factors according to 
\begin{equation}
\frac{i\sum_j   \sum_k\frac{e_{q_k}^2}{3} }{e^2\hat{q}^2}   \lim_{q_4 \rightarrow 0}\partial_{\nu_4}^{q_4}\Bigg\langle \bar{q}(0)\left[\overrightarrow{D}^{\alpha}-\overleftarrow{D}^{\alpha}\right] \gamma^{\beta}  q(0) \Bigg\rangle ^{j, \mu_3,\, \mu_4}_{\overline{\textrm{MS}}(\mu)}
=
\sum_{i=1} ^{6}\omega_{(1)}^{D,i} \, L_i^{\alpha\beta\mu_3\mu_4\nu_4} \, . 
\end{equation}
The decomposition into $\hat{\Pi}_i$ is there identical to the octet case, but their perturbative short-distance quark loop limit is twice the octet case. This means that the for the singlet form factors $\omega _{(1)}^{D,1} $, $\omega _{(1)}^{D,5}$  and $\omega _{(1)}^{D,6}$ , the perturbative limit is instead of Eq.~(\ref{eq:octetpertlimit}) now
\begin{align}\label{eq:singletpertlimit}
    \underline{q_{i}\textrm{ small:}} &
\nonumber \\
\omega _{(1)}^{D,1} 
&
= \frac{-13+12 \log \frac{Q_i}{\mu}}{54\, \pi^2}
\, ,
\nonumber \\
\omega _{(1)}^{D,5} 
&
= \frac{-13+12 \log \frac{Q_i}{\mu}}{54\, \pi^2}
\, ,
\nonumber \\
\omega _{(1)}^{D,6} 
&
= \frac{1}{18\, \pi^2}
\, .
\end{align}
Of course, at higher orders and at the nonperturbative level $\omega_{(1)}^{D,i}\neq 2 \, \omega_{(8)}^{D,i}$.

The main qualitative difference for the singlet piece arises from the gluon operator contributions. There is no $\alpha_s^2$ suppression when $Q_3 \rightarrow \Lambda_{\mathrm{QCD}}$.\footnote{One explicit example is the low-energy theorem that relates $\langle 0|\alpha_s \tilde{G}G | \gamma \gamma \rangle$ to  $\langle 0|\alpha \tilde{F}F | \gamma \gamma \rangle$~\cite{Shifman:1988zk}.} Once again the associated matrix element needs to be form factor decomposed. 
There are in total 7 independent form factors $\omega^{G}_i$ with associated Lorentz structures $L_{i,G}^{\mu_3'\mu_4'\nu_3' \nu_4' \mu_3\mu_4\nu_4}$, so that
\begin{align}
\sum_j    \sum_k\frac{e_{q_k}^{2}}{3} \lim_{q_4 \rightarrow 0}\partial^{\nu_4}_{q_4}\Bigg\langle \frac{1}{2 e^2N_{c}} \, g_s^2 \,  G_{a}^{\nu_3'\mu_3'}G_{a}^{\nu_4'\mu_4'} \Bigg\rangle^{j,\mu_3\mu_4}=\sum_{i=1}^7 \omega^{G}_i L_{i,G}^{\mu_3'\mu_4'\nu_3' \nu_4' \mu_3\mu_4\nu_4} \, .
\end{align}
That only $7$ form factors appears for a matrix element with 7 Lorentz indices follows from substantial symmetry requirements. The ones needed are
\begin{itemize}
    \item Anti-symmetry under $\mu_{3}' \leftrightarrow \nu_{3}'$ exchange.
    \item Anti-symmetry under $\mu_{4}' \leftrightarrow \nu_{4}'$ exchange.
    \item Anti-symmetry under $\mu_{4} \leftrightarrow \nu_{4}$ exchange.
    \item Symmetry under simultaneous exchange of $\mu_{3}'\leftrightarrow \nu_{3}'$ and $\mu_{4}'\leftrightarrow \nu_{4}'$.
    \item Gauge invariance condition under $\mu_3$.
\end{itemize}
The expressions for the Lorentz structures are rather lengthy and have been relegated to the supplementary file {\tt singletbasis.txt}. The decomposition has been arranged in such a way that the analogous matrix element for the photon case saturates to the first form factor, $\omega^{G}_1$, which additionally, for the gluon case, can be linked to $\langle 0|\alpha_s \tilde{G}G | \gamma \gamma \rangle$, whose low-energy limit is known \cite{Shifman:1988zk}. Requiring that there are no kinematic singularities for the $\hat{\Pi}$ leads to
\begin{equation}
\omega^{G}_4=0 \, .
\end{equation}
Employing this condition, we find the $\hat{\Pi}_i$ in the three regions to be
\begin{align}
\underline{q_{3}\textrm{ small:}} &
\nonumber \\
\hat{\Pi}_{1}&=\mathcal{O}\left(\frac{1}{\overline{Q}_{3}^4} \right) \, ,\\ \nonumber
\hat{\Pi}_{4}&=-\frac{16}{3\pi^2 \overline{Q}_{3}^4}
\Big[ \omega_{1}^{G}-3\omega_{2}^{G}+\omega_{3}^{G}-\omega_{5}^{G}-2\omega_{6}^G-2\omega_7^G
\\
&
\qquad
-2\log\frac{\overline{Q}_{3}}{2\mu}\left(\omega_5^G+2\omega_6^G+2\omega_7^G\right) \Big]
\nonumber \\
&+\frac{32}{3\pi^2 \overline{Q}_{3}^4}\left( \omega_5^G+2\omega_6^G+2\omega_7^G \right)\left( 1-\frac{\delta_{12}^2}{Q_3^2}\right) \, ,
\\ \hat{\Pi}_{7}&=\frac{64\, \delta_{12}\, \left(\omega_{6}^G+2\omega_7^G \right)}{3\pi^2 Q_{3}^2 \overline{Q}_{3}^5} \, ,\\
\hat{\Pi}_{17}&=\frac{8}{3\pi^2 Q_3^2\overline{Q}_3^4}\Bigg[\left(-3+8\log\frac{\overline{Q}_{3}}{2\mu}\right) \, \omega_5^G-3\, \omega_6^G +12\, \omega_6^G \, \log\frac{\overline{Q}_{3}}{2\mu}
\nonumber\\
& \qquad
-3\, \omega_7^G+8\, \omega_7^G \, \log\frac{\overline{Q}_{3}}{2\mu} \Bigg] \, ,
\\\hat{\Pi}_{39}&=-\frac{8}{3\pi^2 Q_3^2\overline{Q}_3^4}\Bigg[ \left( -1+4\log\frac{\overline{Q}_{3}}{2\mu}\right) \omega_6^G+\left(-3+8\log\frac{\overline{Q}_{3}}{2\mu}\right)\omega_7^G\Bigg] \, ,\\
\hat{\Pi}_{54}&=\mathcal{O}\left( \frac{1}{\overline{Q}_{3}^5}\right) \, \, ,
\end{align}
\begin{align}
\underline{q_{1}\textrm{ small:}} &
\nonumber \\
\hat{\Pi}_{1}&=\frac{8}{3\overline{Q}_{1}^4\pi^2}\left[\left(1-4\log\frac{\overline{Q}_{1}}{2\mu}\right) \left( \omega_1^G-3\omega_2^G\right)+\left(3-8\log\frac{\overline{Q}_{1}}{2\mu}\right) \omega_3^G\right] \, ,\\
\hat{\Pi}_{4}&=\frac{2}{3\overline{Q}_{1}^2Q_{1}^2\pi^2}\left[-\left(1-4\log\frac{\overline{Q}_{1}}{2\mu}\right)\omega_6^G-\left(3-8\log\frac{\overline{Q}_{1}}{2\mu}\right)\omega_7^G\right] \, ,\\
\hat{\Pi}_{7}&=\mathcal{O}\left(\frac{1}{\overline{Q}_1^4} \right)\, ,\\
\hat{\Pi}_{17}&=\mathcal{O}\left(\frac{1}{\overline{Q}_1^5} \right)\, ,
\\
\hat{\Pi}_{39}&=-\frac{8}{3\overline{Q}_{1}^4 Q_{1}^2\pi^2}\left[-\left(1-4\log\frac{\overline{Q}_{1}}{2\mu}\right)\omega_6^G-\left(3-8\log\frac{\overline{Q}_{1}}{2\mu}\right)\omega_7^G\right]\, ,\\
\hat{\Pi}_{54}&=\frac{8}{3\overline{Q}_{1}^4 Q_{1}^2\pi^2}\left[-\left(1-4\log\frac{\overline{Q}_{1}}{2\mu}\right)\omega_6^G-\left(3-8\log\frac{\overline{Q}_{1}}{2\mu}\right)\omega_7^G\right]\, ,
\end{align}
\begin{align}
\underline{q_{2}\textrm{ small:}} &
\nonumber \\
\hat{\Pi}_{1}&=\frac{8}{3\overline{Q}_{2}^4\pi^2}\left[
\left(1-4\log\frac{\overline{Q}_{2}}{2\mu}\right)\left(\omega_1^G-3\omega_2^G\right)
+\left(3-8\log\frac{\overline{Q}_{2}}{2\mu}\right)\omega_3^G
\right] \, ,\\
\hat{\Pi}_{4}&=\frac{2}{3\overline{Q}_{2}^2Q_{2}^2\pi^2}\left[-\left(1-4\log\frac{\overline{Q}_{2}}{2\mu}\right)\omega_6^G-\left(3-8\log\frac{\overline{Q}_{2}}{2\mu}\right)\omega_7^G\right] \, ,\\
\hat{\Pi}_{7}&=\frac{8}{3\overline{Q}_{2}^4Q_{2}^2\pi^2}\left[-\left(1-4\log\frac{\overline{Q}_{2}}{2\mu}\right)\omega_6^G-\left(3-8\log\frac{\overline{Q}_{2}}{2\mu}\right)\omega_7^G\right] \, ,\\
\hat{\Pi}_{17}&=\mathcal{O}\left(\frac{1}{\overline{Q}_2^5} \right)\, ,
\\
\hat{\Pi}_{39}&=-\frac{8}{3\overline{Q}_{2}^4 Q_{2}^2\pi^2}\left[-\left(1-4\log\frac{\overline{Q}_{2}}{2\mu}\right)\omega_6^G
-\left(3-8\log\frac{\overline{Q}_{2}}{2\mu}\right)\omega_7^G\right]\, ,\\
\hat{\Pi}_{54}&=-\frac{8}{3\overline{Q}_{2}^4 Q_{2}^2\pi^2}\left[-\left(1-4\log\frac{\overline{Q}_{2}}{2\mu})\omega_6^G-(3-8\log\frac{\overline{Q}_{2}}{2\mu}\right)\omega_7^G\right]\, .
\end{align}
Let us note how the $\mu$-dependence on the Wilson coefficients multiplying the gluon matrix elements cancel with the corresponding one of the $D=4$ matrix element involving quark fields. Notice how this is not in contradiction with the fact that no logarithms appeared in the explicit $Q_3 \gg \Lambda_{\mathrm{QCD}}$ matrix elements computed in Sec.~\ref{sec:alphaSLO}, since in that regime the gluonic operator contributions become suppressed by $\alpha^2_{s}(Q_3^2)$ and then they are beyond the computed order.

\section{Conclusions}
\label{sec:conclusions}
In this paper we have studied short-distance constraints for the HLbL in the so-called Melnikov-Vainshtein limit, where two of the photon virtualities are greater than the third, or, $Q_i^2,Q_j^2\gg Q_{k}^2,\Lambda_{\textrm{
QCD}}^2$. The constraints were derived using a systematic and renormalized operator product expansion, through dimension $D=4$ at leading order in $\alpha_s$, thus extending the results of Ref.~\cite{Melnikov:2003xd} and shedding light on the validity of the OPE. We have further shown that in the limiting case $Q_k^2\gg \Lambda _{\mathrm{QCD}}^2 $, our previous results in Refs.~\cite{Bijnens:2019ghy,Bijnens:2020xnl,Bijnens:2021jqo} are reproduced. This completes our fundamental understanding on the validity domains of the quark-gluon picture of the HLbL tensor. 

In the nonperturbative limit without an inherent ordering on $Q_k^2 $ and $\Lambda _{\mathrm{QCD}}^2 $, we decomposed all matrix elements in general form factor decompositions. At dimension $D=3$ we reproduce the known longitudinal and transversal form factors of Ref.~\cite{Melnikov:2003xd}. At dimension $D=4$, we find, to the best of our knowledge, novel relations among Green functions of different dimensionality by using that the $\hat{\Pi}_i$ by construction are free from kinematic singularities. Once this is done, we find that in the octet part all our perturbative ignorance becomes parameterized in terms of three form factors depending on a single energy scale, whose short-distance dependence is known and with a clear definition in terms of Green functions. This opens a clear path to a more reliable determination of the corresponding (subleading) $g-2$ contribution. For the singlet case six additional gluon matrix elements, suppressed by $\alpha_s^2$ in the perturbative regime, emerge, which may limit the associated predictive power.
We leave the nonperturbative completion of the form factors, through model estimates, as well as its impact on $g-2$, for future work. Let us note that the phenomenological impact of the corresponding constraints on $a_\mu^{\textrm{HLbL}}$ has already been studied for the $D=3$ cases in a number of papers~\cite{Colangelo:2019lpu,Colangelo:2019uex,Melnikov:2019xkq,Leutgeb:2019gbz,Ludtke:2020moa,Knecht:2020xyr,Masjuan:2020jsf,Cappiello:2021vzi,Colangelo:2021nkr,Danilkin:2021icn,Leutgeb:2021bpo,Leutgeb:2021mpu,Zanke:2021wiq}, also by implementing them in models.

Let us finally point out that we are currently finalizing work on gluonic corrections to our operator product expansion, which will be the topic of a separate paper. For the dimension $D=3$ contribution, we confirm the statement of Ref.~\cite{Ludtke:2020moa} that it comes with a simple overall coefficient $-\alpha _s /\pi$. At  dimension $D=4$ we have also reproduced the corner expansions in our previous paper Ref.~\cite{Bijnens:2021jqo}.

\section*{Acknowledgements}
J.~B.~is supported by the Swedish
Research Council grants contract numbers 2016-05996 and 2019-03779. 
N.~H.-T. is funded in part by the Albert
Einstein Center for Fundamental Physics at the University of Bern, and in part by the Swedish Research Council, project number 2021-06638. 
A.~R.-S.~is funded in part by MIUR contract number 2017L5W2PT.


\bibliographystyle{JHEP}
\bibliography{refs}

\end{document}